\documentclass[pdftex,epjc3]{svjour3}          % twocolumn

\RequirePackage{graphicx}
\RequirePackage{mathptmx}      % use Times fonts if available on your TeX system
\usepackage{float}
\usepackage{amsfonts}
\usepackage{amsmath}
\usepackage{booktabs}
\RequirePackage{flushend}
\RequirePackage[numbers,sort&compress]{natbib}
\RequirePackage[colorlinks,citecolor=blue,urlcolor=blue,linkcolor=blue]{hyperref}

\begin{document}
	
	\title{An isotropic compact stellar model in curvature coordinate system consistent with observational data}
	
	
	\author{Jitendra Kumar\thanksref{e1,addr1}
		\and
		Puja Bharti\thanksref{e2,addr1}
	}
	
	%\thankstext[$\star$]{t1}{Thanks to the title}
	\thankstext{e1}{e-mail: jitendark@gmail.com}
	\thankstext{e2}{e-mail: pujabharti06@gmail.com}
	\institute{Central University of Jharkhhand, Cheri-Manatu, Ranchi, India.\label{addr1}
	}
	
	\date{Received: date / Accepted: date}

	\maketitle
	
	\section*{Abstract}
	This paper investigates a spherically symmetric compact relativistic body with isotropic pressure profiles  within the framework of general relativity. In order to solve the Einstein's field equations, we have considered the Vaidya–Tikekar type metric potential, which depends upon parameter $K$.  We have presented a charged perfect fluid model, considering $K\notin [0,1]$, which represent compact stars like Her X-1, 4U 1538-52, SAX J1808.4-3658, LMC X-4, SMC X-4,  EXO 1785-248, Cen X-3  and Cyg X-2, to an excellent degree of accuracy. We have investigated the physical features such as the energy conditions, velocity of sound,  surface redshift, adiabatic index of the model in detail and shown that our model obeys all the physical requirements for a realistic stellar model. Using the Tolman-Oppenheimer-Volkoff equations, we have explored the hydrostatic equilibrium and the stability of the compact objects. This model also fulfils the Harrison-Zeldovich-Novikov stability criterion. The results obtained in this paper can be used in analyzing other isotropic compact objects. \\
	\hspace{0.1cm}\\
	\textbf{Key words:} Compact stars; General Relativity; Pressure isotropy; Field equations; Perfect fluid, Exact solutions.
	\newpage
	\section{Introduction}
	General relativity is a great medium for understanding and exploring the gravitational system. Ultra-compact objects like pulsars, neutron stars and black holes have helped scientists to look for exact solutions of the Einstein field equations by modelling physical objects based on observational data, rather than by using mere mathematical excursions. Several theoretical investigations, laboratory experiments and observational tests have been performed during the previous couple of decades. But, it has been difficult to obtain a reliable description of dense compact object. The observational data forming compact stars might be able to provide information about the largest uncertainties in nuclear physics that rely heavily on the equations of state (EoS) at nuclear and supranuclear densities. We might achieve this by estimating their mass and radius which depends on EoS.
	
	The first solution of the Einstein field equations describing a self-gravitating, bounded object was obtained by Schwarzschild \cite{child} about a century ago. The Schwarzschild interior solution describes a uniform density sphere. It was the first approximation in describing the gravitational field of a static, spherically symmetric object. Although this model is not realistic as the propagation speed within the object exceeds the speed of light, the efforts put by Schwarzschild motivated the researchers to search for exact solutions of the Einstein field equations describing self-gravitating objects. As a result, we have a  large number of exact solutions of the field equations describing an outsized number of stellar objects. The analysis of available exact solutions indicates that many of them are unable to describe physically realizable stellar structures \cite{delgaty}. While few of these solutions are only valid in some region of the object, some other solutions display unphysical behaviour in the density and pressure profiles.
	
	A large number of currently existing exact solutions were obtained through various assumptions on the space-time geometry and/or matter content inside the compact object \cite{stephani}. Spherical symmetry is the most common assumption, while modelling static stars. But, there is more freedom in choosing the matter content of the stellar fluid. History tells us that researchers have already worked with perfect fluids, charged interiors, pressure anisotropy, bulk viscosity and scalar fields. Developments in cosmology, inspired the researchers to model stellar structures which includes dark energy, dark matter and phantom energy \cite{benedictis, Rahaman}. 
	
	There is no doubt that an astrophysical structure is not composed of a perfect fluid. However, we may consider relativistic static perfect fluid spheres as first approximation to compact star models. The perfect fluid model necessarily requires that the pressure inside a star should be isotropic, i.e., it should have equal radial $(p_r)$ and tangential $(p_t)$ pressures.
	
	Recent developments in cosmological survey have made us understand the origin and distribution of matter and evolution of compact objects in the Universe. We can measure some of their properties like mass, rotation frequency and emission of radiation. Whereas, measurement of parameters which determines the nature of compact stars is still a great challenge. Properties such as internal composition, mass and radius, which are not directly linked to observations,  requires theoretical models. These theoretical mass and radius are determined by solving the hydrostatic equilibrium equation which convey the equilibrium between gravitational force and pressure. Limited knowledge of nuclear EoS leads to unpredictability of Mass-radius relation. This limits the mass of compact stars. As per Buchdahl \cite{buchdahl}, for a regular fluid sphere with a non increasing mass density, the ratio of its gravitational mass $M$ to that of coordinate radius $R$ satisfies $\frac{M}{R}<\sim \frac{4}{9}$. This constraint arises from the condition that, to prevent gravitational collapse, isotropic pressure does not become infinity at the center of the sphere. In general relativity, the equilibrium of a spherical object is described by the Tolman–Oppenheimer–Volkoff (TOV) equations, and the equation of state is required for its completeness. 
	
	To determine the structure of a compact star, the most common route is to specify an equation of state and then solve the Einstein field equations. Traditionally, this approach has been proved beneficial while using the law of energy conservation in the form of the TOV equation or the equation of hydrodynamical equilibrium.
	
	The equation of state of a compact star is not very clear yet. If one starts with EoS, she generally lands into numerical methods leading to graphical results which lacks in the analysis of local properties of the matter close to the centre of such relativistic stars. Therefore, most researchers prefer to obtain exact solutions of the concerned Einstein’s field equations using ad-hoc methods such  as assuming one of the metric potential. The remaining metric potential is obtained using isotropic conditions for perfect fluid. After this, the physical quantities including pressure, energy density, velocity of sound and adiabatic indices are examined for the reality as well as stability conditions inside the fluid sphere. Examples of some remarkable perfect fluid solutions by assuming metric potential $g_{11}$ can be found in \cite{child, tolman, wyman, patwardhan, buchdahl, kuchowicz1, kuchowicz2, kuchowicz3, bayin, vaidya, finch, mukharjee, gupta1}.
	
	To find the exact solution of the Einstein– Maxwell field equations, Komathiraj and Maharaj \cite{komathiraj1} have used Vaidya and Tikekar ansatz \cite{vaidya} for metric potential with a specified form of electric field intensity. Bijalwan and Gupta \cite{bijalwan1, bijalwan2} have obtained a charged perfect fluid model with generalized electric intensity for all $K \notin (0,1)$. By extending this work, Kumar and Gupta \cite{kumar1, kumar2} obtained another solution for $ K \in (0,1)$. Using this approach, a large number of solutions have been obtained in \cite{patel, sharma, gupta, komathiraj2}. Recently, Kumar et. al. \cite{kumar,kumar3} has obtained perfect fluid charged analogues models with generalized electric intensity for $K \in (0,1)$.  Some other references of solutions using Buchdahl (Vaidya-Tikekar) ansatz (for $K$ in $[0,1]$, $>1$ or $<0$) can be found in \cite{gupta2,prasad2,kumar4,prasad3}.
	
	In this paper, we are going to use a physically viable Vaidya and Tikekar \cite{vaidya} metric potential to obtain a closed-form solution of the Einstein field equations for a spherically symmetric isotropic matter distribution. We will use this solution to develop feasible models for compact stars with some standard observed mass and radius as proposed in \cite{prasad, kumar3}. To find out the model parameters, we will utilize the boundary conditions, which says that interior spacetime metric matches the exterior Schwarzschild metric at surface and radial pressure is zero across the boundary. Due to the complexity of the solution, we will use graphical approach to verify if the matter variables of the model satisfy criteria for realistic star. 
	
	This paper has been organized as mentioned below:\\
	In Sect. 2, the Einstein field equations for the isotropic system of the compact object has been presented. In Sect. 3, by assuming the Vaidya-Tikekar metric potential, the relevant field equations has been solved to develop a new model. In Sect. 4, an analytical and graphical representations has been performed to check the physical acceptability and stability of the model. For this we have used recent measurements of mass and radius of stars SMC X-1, Her X-1, 4U 1538-52, SAX J1808.4-3658, LMC X-4, EXO 1785-248 and Cyg X-2. Finally, Sect. 5 is devoted to conclusion.
	
	\section{Metric and the Field equations}
	Let us consider the line element to describe the static and  spherically symmetric stellar system  in curvature coordinates $(x^i) = (t,r,\theta,\phi)$\\
	\begin{equation}
		ds^2=e^{\nu(r)}dt^2-e^{\lambda(r)}dr^2-r^2(d\theta^2+sin^2\theta d\phi^2),
		\label{metric}
	\end{equation}
	where the metric potentials $\nu(r)$ and $\lambda(r)$ are arbitrary functions of radial coordinate $r$. These potentials plays a key role in determining the surface redshift and gravitational mass function respectively. \\
	The Einstein-Maxwell field equations for obtaining the hydrostatic stellar structure of the charged sphere can be written as 
	\begin{equation}
		-\kappa(T_{j}^{i}+E_{j}^{i})=R_{j}^{i}-\frac{1}{2}R\delta_{j}^{i}=G_{j}^{i},
		\label{ife}
	\end{equation} 
	where $\kappa= \frac{8\pi G}{c^4}$,  $G$ here stands for gravitational constant and $c$ is the speed of light, $R_{j}^{i}$ and $R$ represent Ricci Tensor and Ricci Scalar respectively. Throughout the discussion we will take $G=c=1$, as geometrized units. 
	Since we are assuming that matter within the star is a charged perfect fluid, the corresponding energy-momentum tensor $T_{j}^{i}$ and electromagnetic field tensor $ E_{j}^{i}$ will be
	\begin{equation}
		T_{j}^{i} = (\rho + p) v^{i} v_{j }-p\delta_{j}^{i} 
	\end{equation}
	and
	\begin{equation}
		E_{j}^{i} = \frac{1}{4\pi}(-F^{im}F_{jm}+\frac{1}{4}F^{mn}F_{mn}),
	\end{equation}
	where, $\rho(r)$ is the energy density, $p(r)$ is the isotropic pressure, $F_{ij}$ is anti-symmetric electromagnetic field strength tensor defined as
	$F_{ij} = \frac{\partial A_{j}}{\partial x_{i}}-\frac{\partial A_{i}}{\partial x_{j}}$ which satisfies Maxwells equations, 
	\begin{center}
		$F_{ik,j}+F_{kj,i}+F_{ji,k}=0$ and $[\sqrt{-g} F^{ik}]_{,k}=4\pi J^{i}\sqrt{-g}$
	\end{center}
	Here $A_j = (\phi(r), 0, 0, 0)$ is the potential and $J^{i}$ is the electromagnetic current vector defined as $J^i = \frac {\sigma}{\sqrt{g_{44}}}\frac {dx^i}{dx^4}= \sigma \nu^i$, where $\sigma = e^{(\nu/2)}J^0$ represents the charge density, $g$ is the determinant of the metric $g_{ij}$ which is defined by $g=-e^{\nu+\lambda} r^4 sin^{2}\theta$ and $J^0$ is the only non-vanishing component of the electromagnetic current $J^i$ for the static spherically symmetric stellar system. Since the field is static, we have $\nu=(0,0,0,\frac{1}{\sqrt{g_{44}}})$.\\
	Also, the total charge within a sphere of radius $r$ is given by 
	\begin{equation}
		q(r) = r^2E(r) = 4\pi \int_{0}^r J^0 r^2 e^{(\nu+\lambda)/2}dr,
		\label{dense}
	\end{equation}
	where, $E(r)$ is the intensity of the electric field. \\
	Thus, for the spherically symmetric metric of Eq. (\ref{metric}) the Einstein-Maxwell field equation (\ref{ife}) provides the following relationship \cite{landau}:
	\begin{equation}
		\frac{\lambda'}{r}e^{-\lambda}+\frac{1-e^{-\lambda}}{r^2}=c^2 \kappa \rho + \frac{q^2}{r^4} \ ,\
		\label{fe1}
	\end{equation}
	\begin{equation}
		\frac{\nu'}{r}e^{-\lambda}-\frac{1-e^{-\lambda}}{r^2}=\kappa p- \frac{q^2}{r^4} \ ,\
		\label{fe2}
	\end{equation}
	\begin{equation}
		\Big(\frac{\nu''}{2}-\frac{\lambda' \nu'}{4}+\frac{\nu'^2}{4}+\frac{\nu'-\lambda'}{2r}\Big)e^{-\lambda}=\kappa p+ \frac{q^2}{r^4} 
		\label{fe3}
	\end{equation}
	Here prime denotes differentiation with respect to $r$. Using Eqs. (\ref{fe2}) and (\ref{fe3}), we can obtain 
	\begin{equation}
		\Big(\frac{\nu''}{2}-\frac{\lambda' \nu'}{4}+\frac{\nu'^2}{4}-\frac{\nu'+\lambda'}{2r}-\frac{1}{r^2}\Big)e^{-\lambda}+\frac{1}{r^2}= \frac{2q^2}{r^4} 
		\label{fe4}
	\end{equation}
	We can get the definition of charged density $\sigma$ by substituting this value in eq. (\ref{dense}).\\ \\
	Let us consider $m(r)$ to be the mass function for an electrically charged fluid sphere, given as \cite{florides}
	\begin{equation}
		m(r)=\frac{\kappa}{2}\int_{0}^{r}\left(c^2\rho r^2+r \sigma q e^{\lambda/2}\right)dr=\frac{r}{2}\left(1-e^{-\lambda}+\frac{q^2}{r^2}\right).
		\label{mass}
	\end{equation}
	Consider $r = R$ as the outer boundary of the fluid sphere. The unique exterior metric for a spherically symmetric charged distribution of matter is the  Reissner-N\"{o}rdstro metric
	\begin{equation}
		ds^2 =  \Big(1-\frac{2M}{r}+\frac{Q^2}{r^2}\Big)dt^2 - \Big(1- \frac{2M}{r}+\frac{Q^2}{r^2}\Big)^{-1}dr^2-r^2\big(d\theta^2 + sin^2\theta d\phi^2\big)
		\label{extmetric}
	\end{equation}
	where, $ M= m( R)$, is total gravitational mass and $Q=q(R)$ is total electric charge.
	\section{Exact solutions of the models for isotropic stars}
	We are going to uncover the solutions to Einstein's field equations for isotropic fluid matter. To achieve this, we have to solve 3 equations (\ref{fe1},\ref{fe2}\& \ref{fe4}) for 5 unknown functions. Let's specify two variables a priori to solve these equations analytically.\\\\  
	Let's consider the metric ansatz, given by Vaidya and Tikekar \cite{vaidya}
	\begin{equation}
		e^\lambda=\frac{K(1+Cr^2)}{K+Cr^2},
		\label{lambda}
	\end{equation}
	and a new variable as
	\begin{equation}
		e^\nu=Z^2 (r)
		\label{nu}
	\end{equation}
	where $C$ and $K$ are some constant parameters. This choice of metric potential provides a singularity free solution at $r = 0$ and $e^\lambda(0) = 1$. Vaidya and Tikekar \cite{vaidya} had  considered this metric potential to study spheroidal spacetimes governing the behavior of superdense stars. Several works utilizing this form of metric potential can be found in literature \cite{korkina,kumar,prasad}.\\\\
	In order to get the exact solutions more efficiently, we will use the above substitutions, so that we can transform the field equations to an equivalent form as, 
	\begin{equation}
		c^2\kappa \rho+\frac{q^2}{r^4}=\frac{C(K-1)(3+Cr^2)}{K(1+Cr^2)^2}
		\label{fe5}
		\end {equation}
		\begin{equation}
			\kappa p-\frac{q^2}{r^4}=\frac{K+Cr^2}{K(1+Cr^2)} \frac{2Z'}{rZ}+\frac{C(1-K)}{K(1+Cr^2)}
			\label{fe6}
		\end{equation}
		and
		\begin{equation}
			\frac{d^2Z}{dr^2}-\Big[\frac{K+2KCr^2+C^2r^4}{r(1+Cr^2)(K+Cr^2)}\Big]\frac{dZ}{dr}+ \Big[\frac{C^2r^2(K-1)}{(K+Cr^2)(1+Cr^2)}-\frac{2Kq^2(1+Cr^2)}{r^4(K+Cr^2)}\Big]Z=0
			\label{fe7}
		\end{equation}
		Here we are considering the charged perfect fluid distribution represented by metric (\ref{metric}) when $K$ $\notin$ $[0,1]$, i.e., for $K<0$ and $K>1$.\\\\
		To get a convinient form of the above equations	let's introduce the transformation 
		\begin{equation}
			X=\sqrt{\frac{K+Cr^2}{K-1}}
			\label{X}
		\end{equation}
		where, $0<C<\frac{|K|}{R^2}$ is a parameter, which characterizes the geometry of star. \\
		Substituting the value of $X$ into eq. (\ref{fe7}), we get,
		\begin{equation}
			\frac{d^2Z}{dX^2}-\frac{X}{X^2-1}\frac{dZ}{dX}+(K-1)\Big[\frac{1}{X^2-1}-\frac{2K(1+Cr^2)q^2}{C^2r^6} \Big]Z
			\label{fe71}
		\end{equation}
		It is obvious from eq. (\ref{X}) that when $K$ is negetive $X$ is less than 1 and when $K>1$, we get $X>1$. \\
		Let's use the transformation
		\begin{equation}
			Z=(1-X^2)^{1/4}Y \ \  when \ K<0 \ \ \& \ \	Z=(X^2-1)^{1/4}Y \ \  when \ K>1  	 
			\label{z}
		\end{equation}
		to convert eq. (\ref{fe71}) into the normal form
		\begin{equation}
			\frac{d^2Y}{dX^2}+\phi Y=0,
			\label{nf1}
		\end{equation}
		where, 
		\begin{equation}
			\phi= \frac{1}{1-X^2}\Big[ 1-K+\frac{2Kq^2(1+Cr^2)^2}{C^2r^6}+\frac{3X^2+2}{4(X^2-1)}\Big]
			\label{nf2}
		\end{equation}
		It is difficult to solve the second order differential equation (\ref{nf1}) using standard techniques. In order to solve this differential equation, let's take
		\begin{equation}
			\phi= -\frac{2a_1}{X^2(a_1+a_2X)}
			\label{phi}
		\end{equation}
		where, $a_1, a_2 \in \mathbb{R}$ such that $a_1$ is non-zero. \\
		We have made this choice for $\phi$, as it will later become evident that it simplifies the analysis. For the stars which we have considered here, such a choice gives physically viable electric field intensity.\\
		Putting this value of $\phi$ from eq. (\ref{phi}) to eq. (\ref{nf1}), the resulting differential equation becomes
		\begin{equation}
			X^2(a_1+a_2X)\frac{d^2Y}{dX^2}-2a_1Y=0
			\label{nf4}
		\end{equation}
		It's solution can be given by 
		\begin{equation}
			Y=\frac{a_1+a_2X}{X} \Big [ A_1\frac{a_1}{a_2^3}S(X)+A_2\Big ]
			\label{soly}
		\end{equation}
		where, $A_1$ and $A_2$ are arbitrary constants and
		\begin{equation}
			S(X)=\frac{\sec^2\Big(\tan^{-1}\sqrt{\frac{a_2X}{a_1}}\Big)}{2}-\frac{\cos^2\Big(\tan^{-1}\sqrt{\frac{a_2X}{a_1}}\Big)}{2}+2\log \mid \cos\Big(\tan^{-1}\sqrt{\frac{a_2X}{a_1}}\Big)\mid 
		\end{equation}
		
		Using eqs. (\ref{z}) and (\ref{soly}) we get the value of $Z$ as,
		\begin{eqnarray}
			\nonumber	Z= A_1(1-X^2)^{1/4}\frac{a_1+a_2X}{X} \Big [\frac{a_1}{a_2^3}S(X)+\frac{A_2}{A_1}\Big ], \ \ \text{when} \ \ K<0  \\
			Z= A_1(X^2-1)^{1/4}\frac{a_1+a_2X}{X} \Big [\frac{a_1}{a_2^3}S(X)+\frac{A_2}{A_1}\Big ], \ \ \text{when} \ \ K>1
			\label{solz}
		\end{eqnarray}
		We will show in the next section that the obtained metric function $Z = e^{\nu/2}$ is finite, free from singularity at centre with $\nu'(0)=0$ and monotonically increasing throughout the stellar interior, i.e., it satisfies the prerequisites for any physically acceptable model provided by  Lake \cite{lake}. This function will act as the second necessary condition which we have imposed to generate the model along with the assumption (\ref{lambda}).\\
		Now, let's obtain the expression for electric charge, energy density and pressure.\\\\
		On comparing eqs. (\ref{nf2}) and (\ref{phi}), we get the definition of electric field intensity as
		\begin{equation}
			E^2=\frac{q^2}{r^4}= \frac{C^2r^2}{2K(1+Cr^2)^2}\Big[ \frac{5}{4}\frac{1}{(1-X^2)}-\frac{2a_1}{X^2(a_1+a_2X)}(1-X^2)+K-\frac{7}{4}\Big]
			\label{charge}
		\end{equation}
	
		Putting eqs. (\ref{charge}) and (\ref{solz}) into eqs. (\ref{fe5}) and (\ref{fe6}) respectively, we obtain the expressions for energy density and pressure as:
		\begin{equation}
			c^2\kappa \rho=\frac{C(K-1)(3+Cr^2)}{K(1+Cr^2)^2}- \frac{C^2r^2}{2K(1+Cr^2)^2}\Big[\frac{5}{4(1-X^2)}-\frac{2a_1(1-X^2)}{X^2(a_1+a_2X)}+K-\frac{7}{4}\Big]
			\label{density}
			\end{equation}
			\begin{align}
				\kappa p=\frac{CX^2}{K(X^2-1)} \Big[\frac{P_1 P_2+P_3 P_4}{P_2P_5}\Big]-\frac{C}{K(X^2-1)}+ \frac{C^2r^2}{2K(1+Cr^2)^2}\Big[\frac{5}{4(1-X^2)}-\frac{2a_1(1-X^2)}{X^2(a_1+a_2X)}+K-\frac{7}{4}\Big]
				\label{pressure}
			\end{align}
			
			From eqs. (\ref{density}) and (\ref{pressure}), graidiant of density and pressure can be obtained as,
			\begin{equation}
				c^2\kappa \frac{d\rho}{dr}=C^2r\Big[ D_6-D_7-D_8 \Big]
				\label{grad.d}
				\end{equation}
				\begin{equation}
					\kappa \frac{dp}{dr}=C^2r\Big[\frac{X^2}{K(X^2-1)} \frac{D}{P_2 P_5}+\frac{2}{K(1-K)(X^2-1)^2}\Big(\frac{P_1 P_2+P_3 P_4}{P_2 P_5}-1\Big)+D_7+D_8 \Big]
					\label{grad.p}
					\end{equation}
					respectively.\\ 
					\textbf{See Appendix A} for values of $P_i \ (1\le i \le 5)$, $D$ and $D_j \ (1\le j \le 8)$.		
					\section{Physical features and stability analysis of the model}
					In this section, we are going to perform some analytical calculations to ensure that this model is obeying essential physics for a stellar structure throughout the interior and outer surface. We will do the stability analysis of the model by studying general physical properties and plotting several figures for some of the compact star candidates. The solutions found in this paper might be useful in study of relativistic compact stellar objects.
					\subsection{\textbf{Boundary Conditions}}
					To explore the boundary conditions, we are going to use the fact that all astrophysical objects are immersed in vacuum or almost vacuum space-time. Also, the interior metric (\ref{metric}) joins smoothly at the surface of spheres ($r=R$) to the exterior metric (\ref{extmetric}). In order to match smoothly on the boundary surface $r = R$, we will impose the boundary conditions which are equivalent to the following two conditions: 
					\begin{equation}
						e^{\nu(R)}=Z^2(R)=1- \frac{2M}{R}+\frac{Q^2}{R^2}, \ \& \ e^{-\lambda(R)} =1- \frac{2M}{R}+\frac{Q^2}{R^2},	
						\label{bcon1}
					\end{equation}
					and  \begin{equation}
						P(R)=0	
						\label{bcon2}
					\end{equation}
					where,  $ Q=q(R)$. 
					Using these boundary conditions (\ref{bcon1},\ref{bcon2}), we can easily obtain the constants $A_1$ and $A_2$ (see Appendix B).
					
					For a given radius $R$, we can determine the total mass $M$ of the star and vice-versa. Keeping in mind the constraints on the mass-radius ratio $\Big(\frac{2M}{R}\le \frac{8}{9}\Big)$ \cite{buchdahl, bondi}, we have demonstrated that for some particular values of the parameters, we can generate specific mass and radius of some well known pulsars. In this process we have used true values of $c$ and $G$ at appropriate places. Some of such possibilities are tabulated in Table \ref{t1}.
					\begin{table}[htbp]
						\caption{The approximate values of the masses $M$, radii $R$, and the constants $a_1$, $a_2$, $C$ and $K$ for the compact stars}
						\label{t1}
						\begin{tabular}{lccccccc}
							\toprule
							\textbf{Compact Star} & $a_1$  & $a_2$ &$C(/km^2)$ & $K$  &$M/M_\odot$&$R (km)$&$M/R$ \\ 
							\midrule
							Her X-1 \cite{prasad}&0.7&2.6&0.003078799&-1.313519 &0.85&8.1  & 0.15475 \\
							4U 1538-52 \cite{prasad}&0.07&4.6 &0.003393997&-1.208442&0.87&7.866& 0.16314  \\
							SAX J1808.4-3658 \cite{prasad}&0.07&4.58&0.003369272&-1.18123&0.9 &7.951	& 0.16696 \\
							LMC X-4 \cite{kumar3}&0.07&2.6   &0.003192728&-1.02362&1.04&8.301& 0.18479 \\
							SMC X-4 \cite{prasad}&0.07&2.6  &0.003077453&-0.944478 &1.29&8.831& 0.21546 \\
							EXO 1785-248 \cite{prasad}&0.7&2.6&0.003448064&-1.132761 &1.3 &8.849& 0.21669 \\
							Cen X-3 \cite{kumar3}&0.5&1.2&0.003561436&-1.150118&1.49 &9.178& 0.23945 \\
						Cyg X-2 \cite{kumar3}&0.07   &4.1  &4.213965167&2.106839 &1.71&8.313& 0.303519 \\
							\bottomrule 
						\end{tabular}
					\end{table}
					\begin{table}[H]
						\centering
						\caption{Numerical values of surface charge ($q_s$), central density ($\rho_0$), surface density ($\rho_s$), central pressure ($p_0$) and surface redshift ($z_s$) of compact star candidates.}
						\label{t2}
						\begin{tabular}{lccccc}
							\toprule
							Compact star &$q_s (C)$& $\rho_0 (g/cc)$ & $\rho_s (g/cc)$ & $p_0 (Pa)$& $z$  \\
							\midrule 
							Her X-1 &$1.10029\times 10^{20}$&$8.72888\times 10^{14}$&$6.33727\times 10^{14}$&$4.26198\times 10^{33}$&$0.191824$  \\ 
							4U 1538-52&$8.75131\times 10^{19}$&$9.9841\times 10^{14}$&$7.21771\times 10^{14}$&$7.40078\times 10^{33}$&$0.210164$ \\ 
							SAX J1808.4-3658 &$9.11424\times 10^{19}$&$1.00147\times 10^{15}$&$7.20756\times 10^{14}$&$7.62543\times 10^{33}$&$0.21649$ \\ 
							LMC X-4&$1.10765\times 10^{20}$&$1.01599\times 10^{15}$&$7.22473\times 10^{14}$&$8.53183\times 10^{33}$&$0.246591$ \\ 
							SMC X-4 &$1.42547\times 10^{20}$ &$1.06235\times 10^{15}$&$7.23379\times 10^{14}$&$1.09311\times 10^{34}$&$0.303832$  \\ 
							EXO 1785-248&$1.87705\times 10^{20}$&$1.04499\times 10^{15}$&$6.83534\times 10^{14}$&$6.32214\times 10^{33}$&$0.291296$ \\ 
						Cen X-3&$2.33211\times 10^{20}$&$1.07173\times 10^{15}$&$6.67313\times 10^{14}$&$5.29925\times 10^{33}$&$0.326182$ \\ 
							Cyg X-2&$2.79936\times 10^{20}$&$3.56354\times 10^{17}$&$3.44725\times 10^{14}$&$9.36725\times 10^{36}$&$0.448752$ \\ 
							\bottomrule
						\end{tabular}
					\end{table}	
					\subsection{\textbf{Regularity and Reality Conditions}}
\begin{figure}[H]
							\includegraphics[width=8cm]{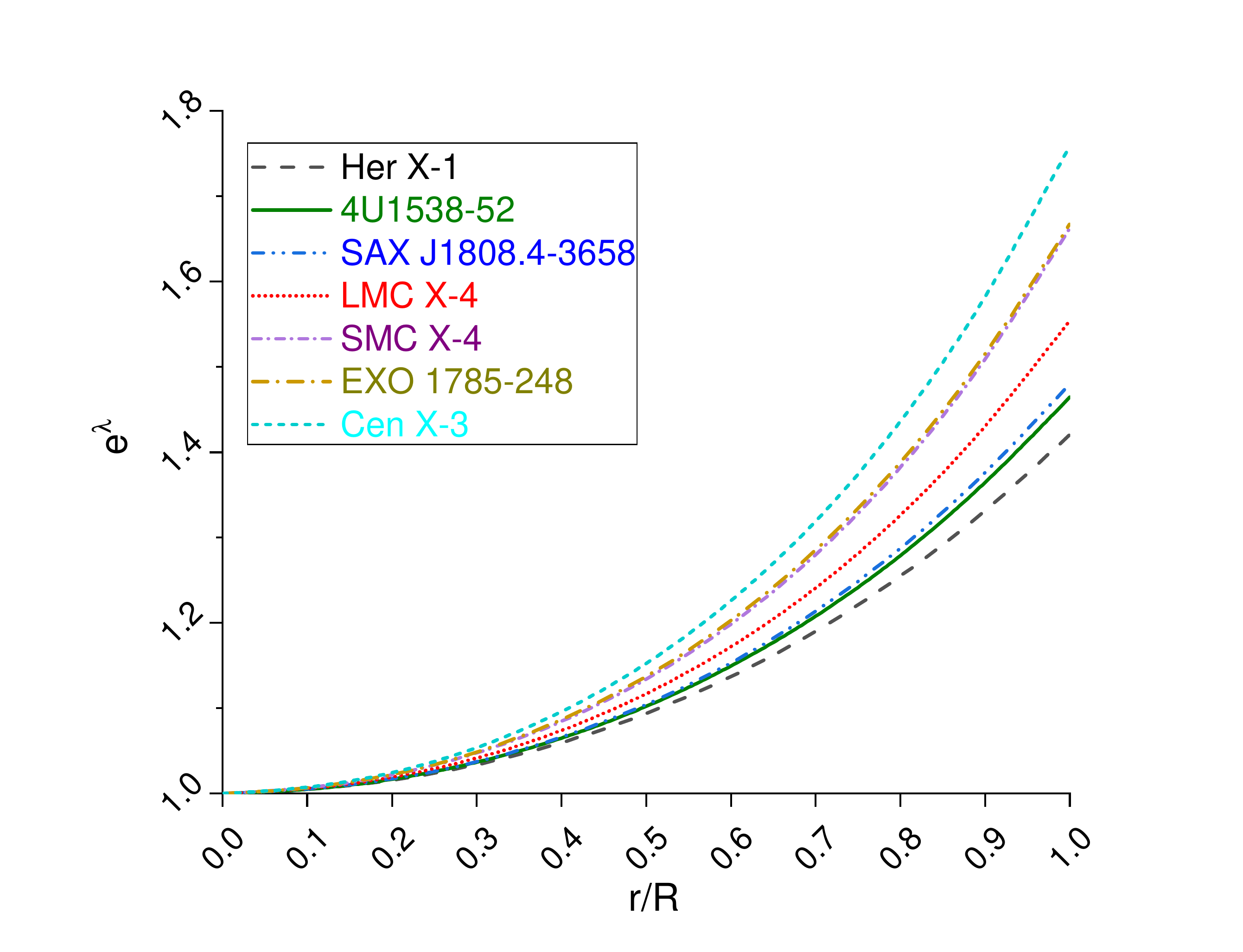}\includegraphics[width=8cm]{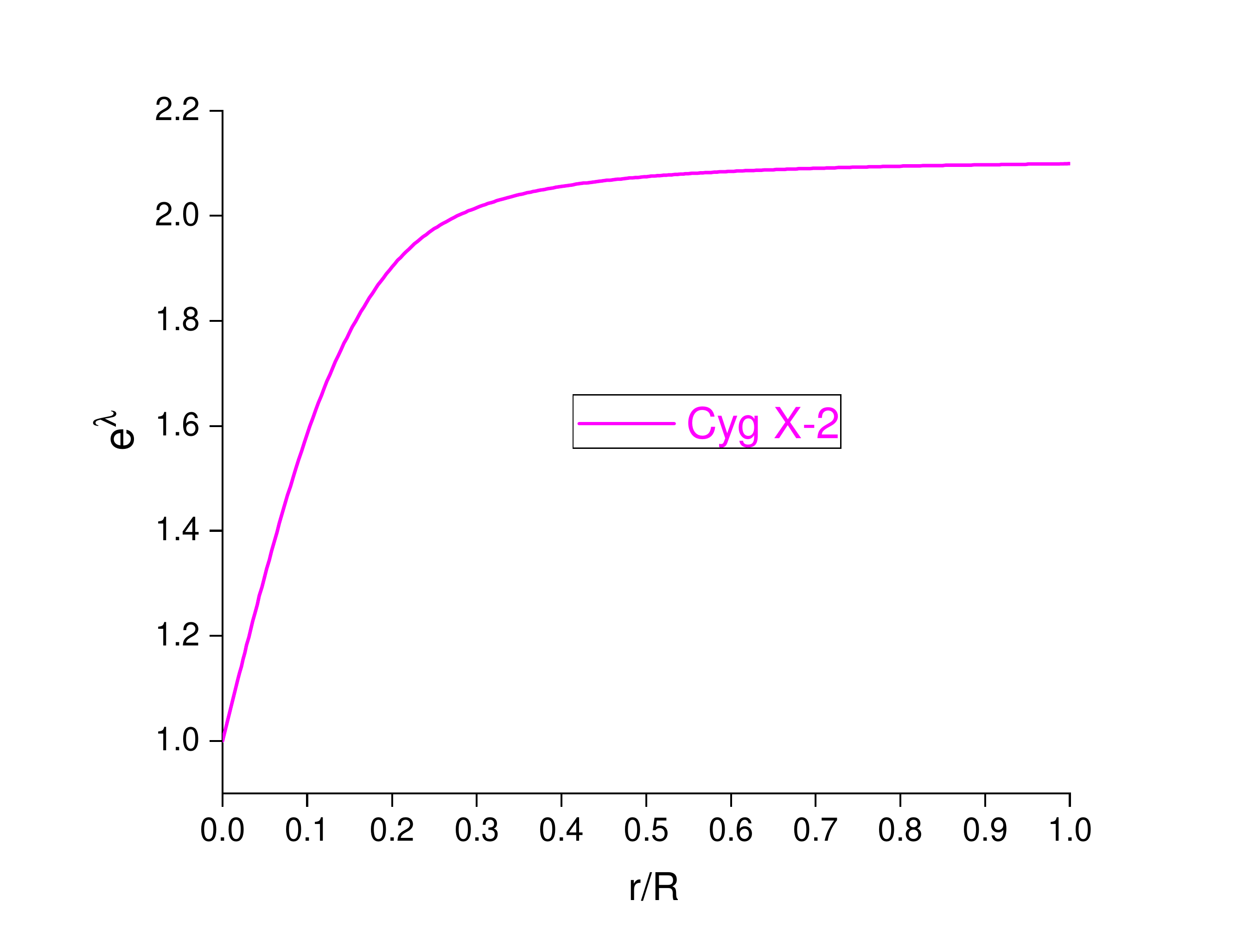}
							\caption{Behavior of $e^\lambda$ within the stellar configuration of star candidates Her X-1, 4U 1538-52, SAX J1808.4-3658, LMC X-4, SMC X-4,  EXO 1785-248, Cen X-3  ($K<0$) and Cyg X-2 ($K>1$).}\label{f1}
					\end{figure}

\begin{figure}[H]
							\includegraphics[width=8cm]{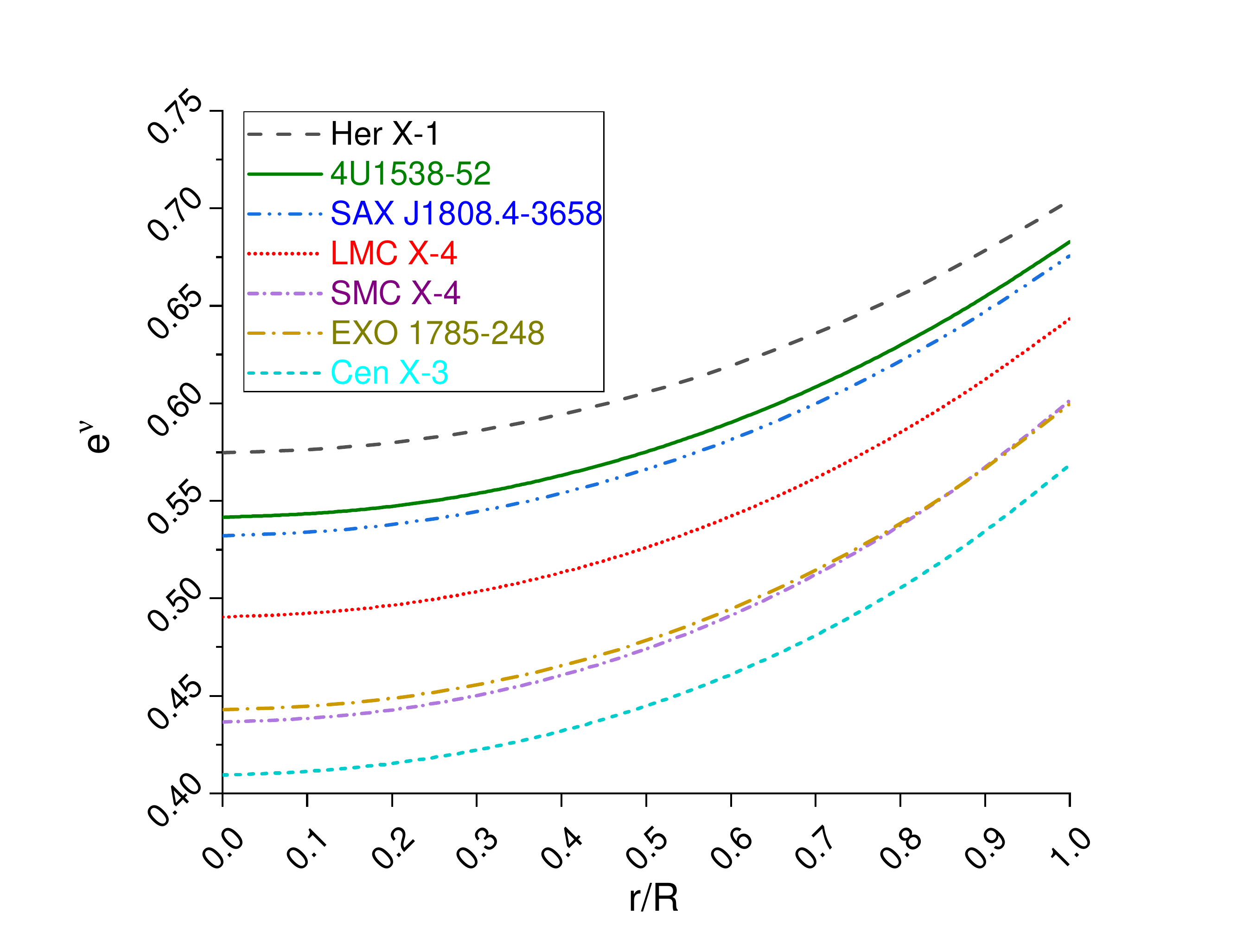}\includegraphics[width=8cm]{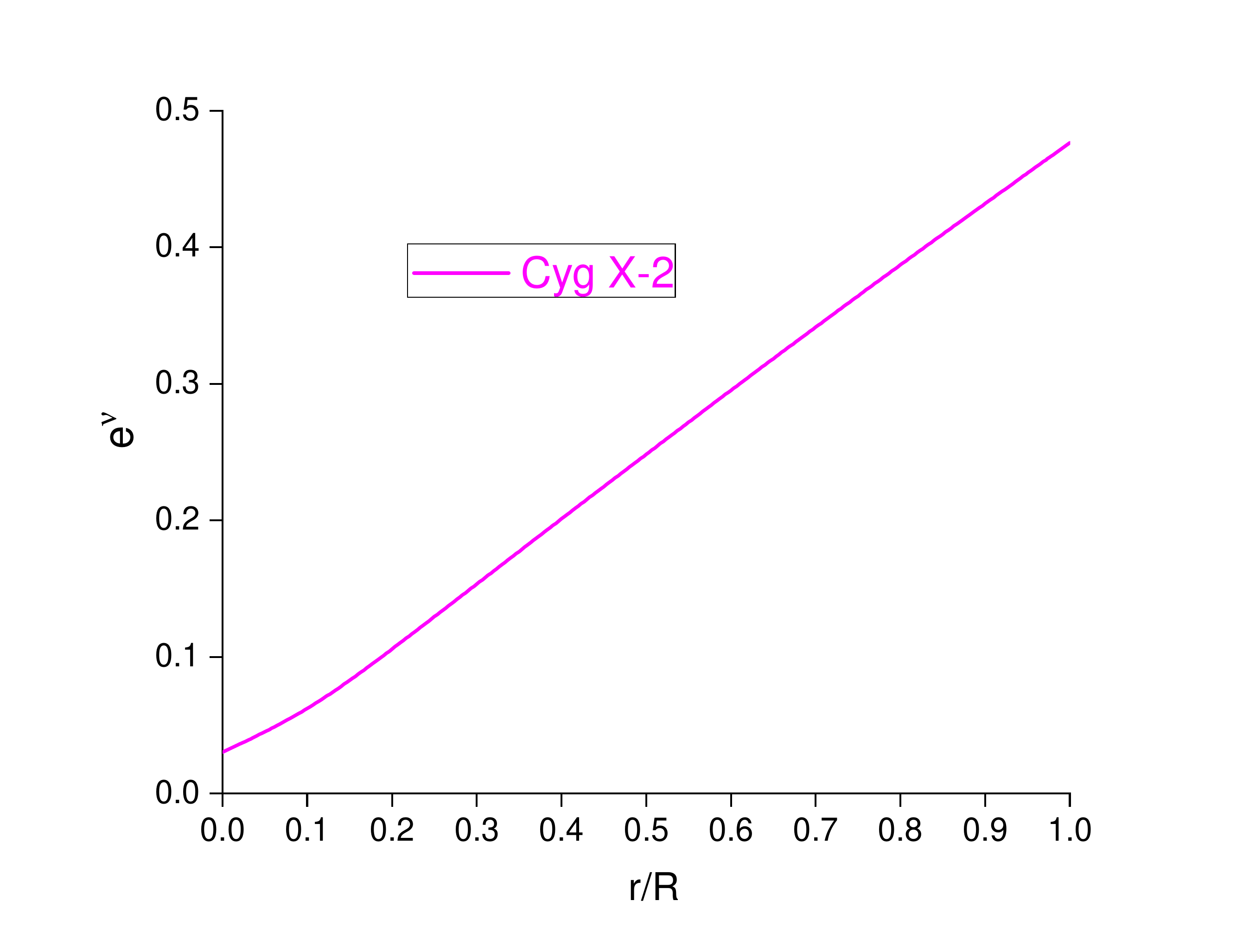}
							\caption{Behavior of $e^\nu$ within the stellar configuration of star candidates Her X-1, 4U 1538-52, SAX J1808.4-3658, LMC X-4, SMC X-4,  EXO 1785-248, Cen X-3  ($K<0$) and Cyg X-2 ($K>1$).}\label{f2}
					\end{figure}

					It is clear from fig (\ref{f1}) and (\ref{f2}) that the obtained metric potentials $e^\lambda$ and $e^\nu$ are free from physical and geometrical singularities. Additionally, they are finite and monotonically increasing throughout the stellar interior. Thus, the behavior of metric functions is consistent with the requirements.\\ 
					For physical feasibility of the model it is also required that\\
					• the energy density is positive definite and its gradient is negative everywhere within the radius.\\
					• for an isotropic fluid distribution  pressure is positive definite and the pressure gradient is negative within the stellar interior.\\ 
					\begin{figure}
							\includegraphics[width=8cm]{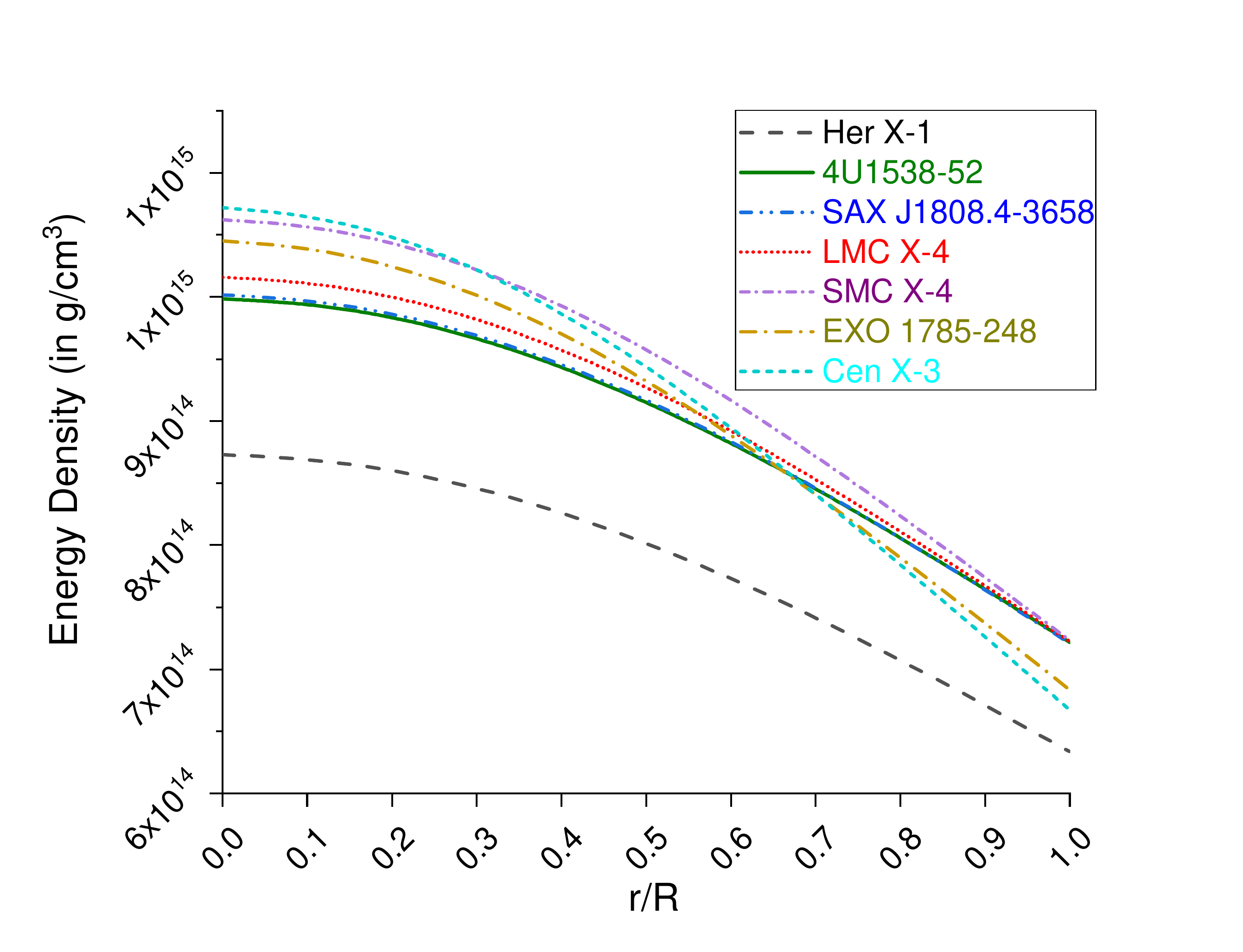}\includegraphics[width=8cm]{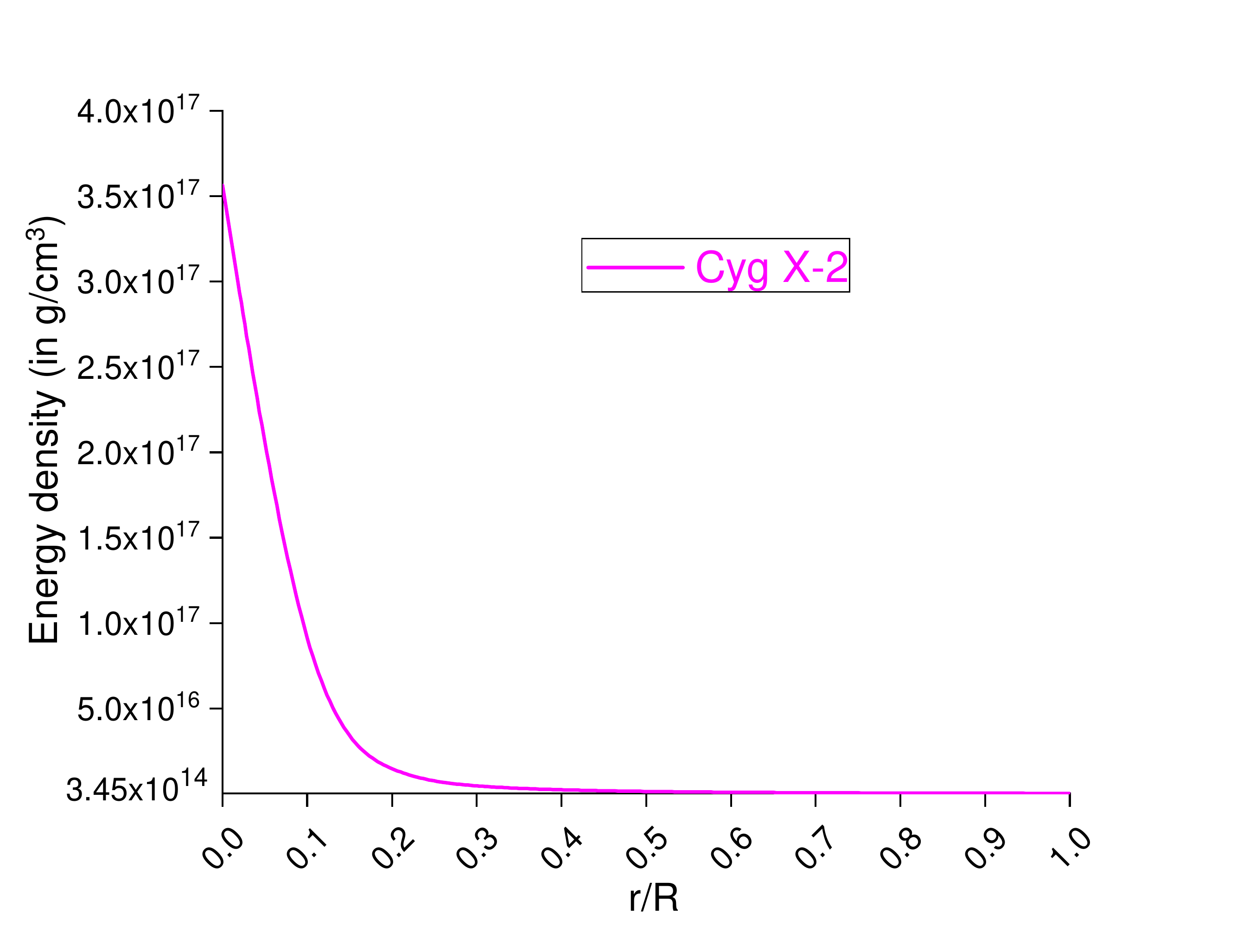}
							\caption{Variation of density with respect to fractional radius (r/R) for star candidates Her X-1, 4U 1538-52, SAX J1808.4-3658, LMC X-4, SMC X-4,  EXO 1785-248, Cen X-3  ($K<0$) and Cyg X-2 ($K>1$).}\label{d}
					\end{figure}
					\begin{figure}
							\includegraphics[width=8cm]{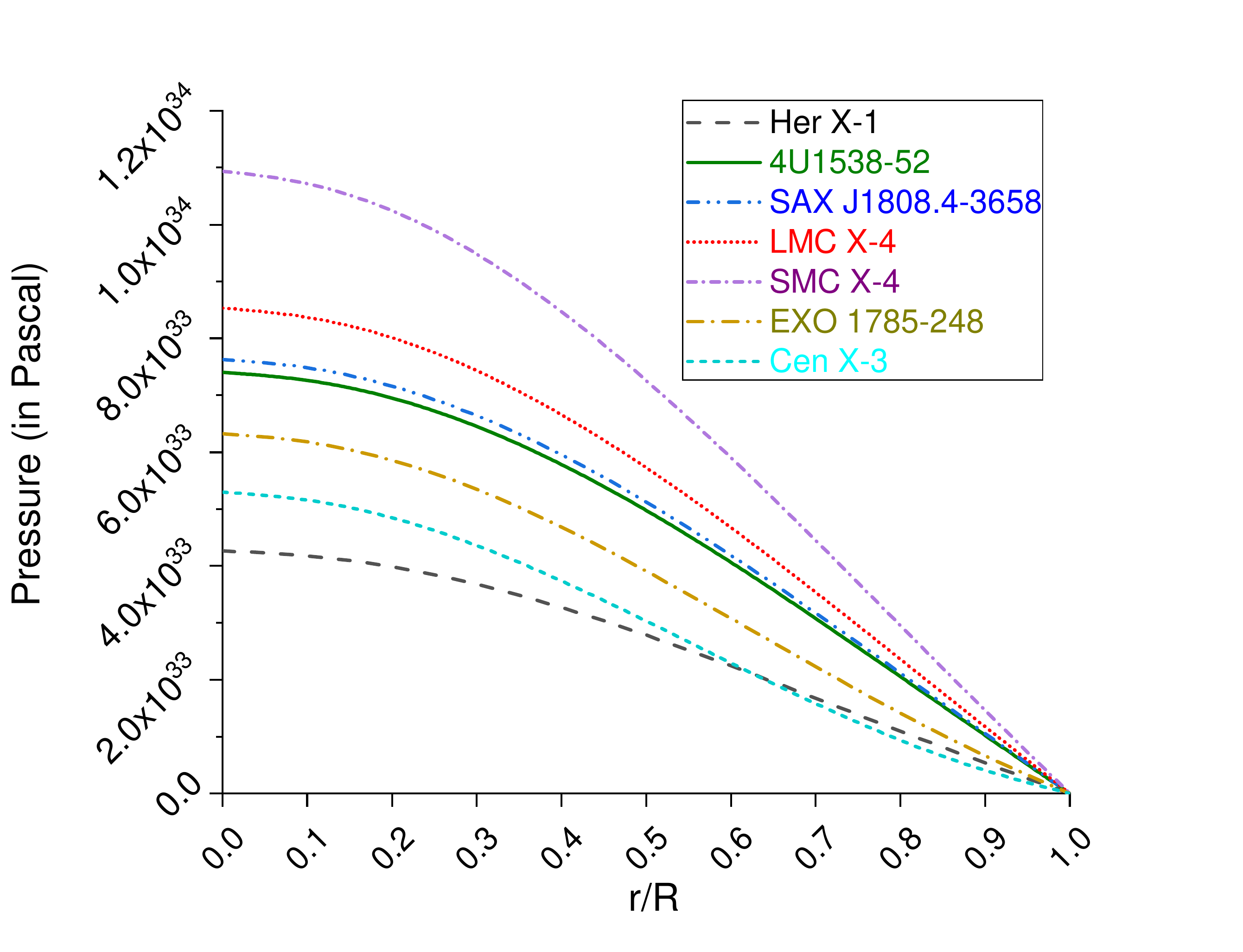}\includegraphics[width=8cm]{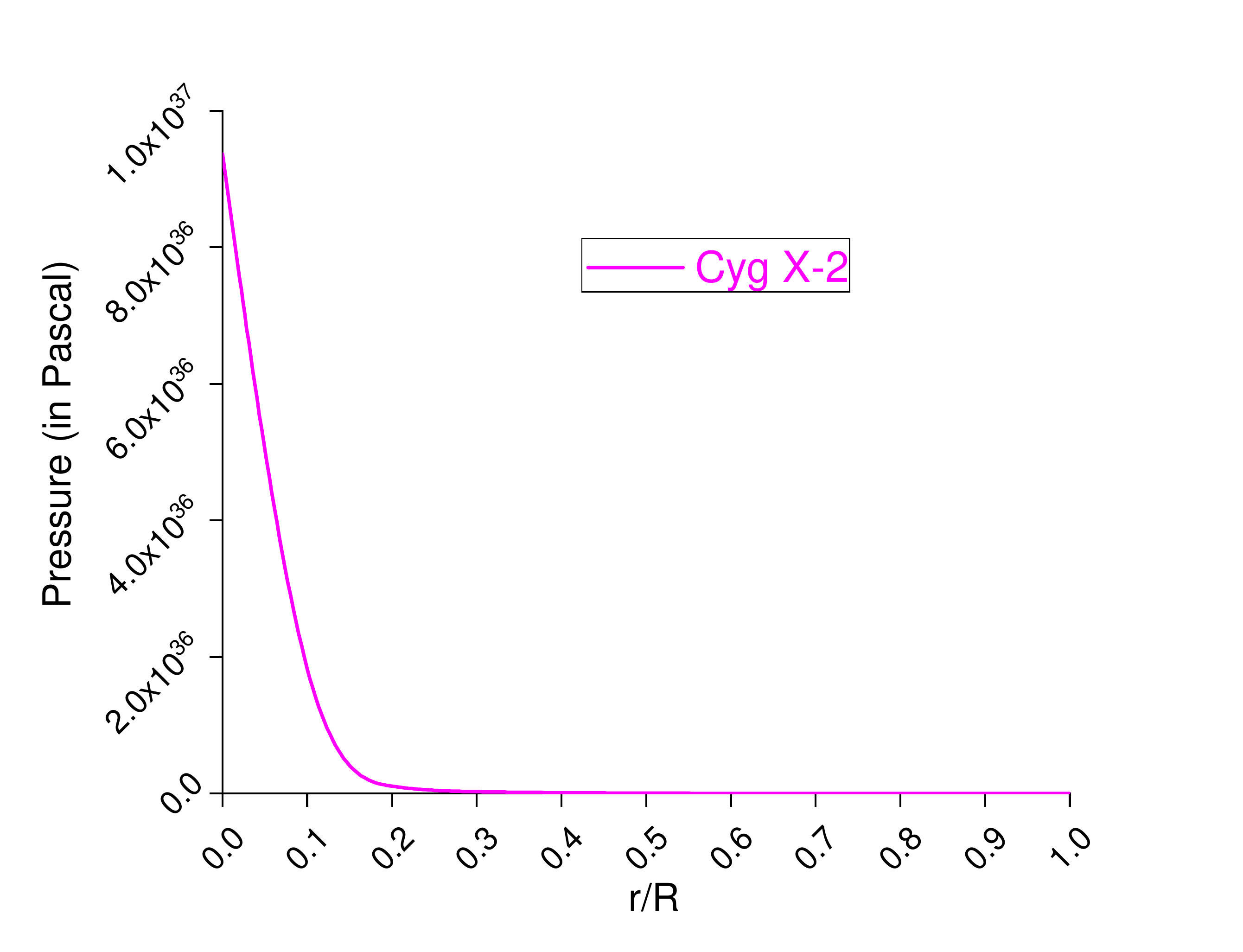}
							\caption{Variation of pressure with respect to fractional radius (r/R) for star candidates Her X-1, 4U 1538-52, SAX J1808.4-3658, LMC X-4, SMC X-4,  EXO 1785-248, Cen X-3  ($K<0$) and Cyg X-2 ($K>1$).}\label{p}
					\end{figure}
					Graphs in Fig. (\ref{d}) and (\ref{p}) indicate that the energy density is positive with a maximum value at the centre and the pressure is finite and vanishes at the boundaries for each considered star candidates. Also, both pressure as well as density are monotonically decreasing in nature towards the surface of star. We have taken the same values of the constants as mentioned in Table \ref{t1}.
						\subsection{\textbf{Electric charge}}
						\begin{figure}[H]
								\includegraphics[width=8cm]{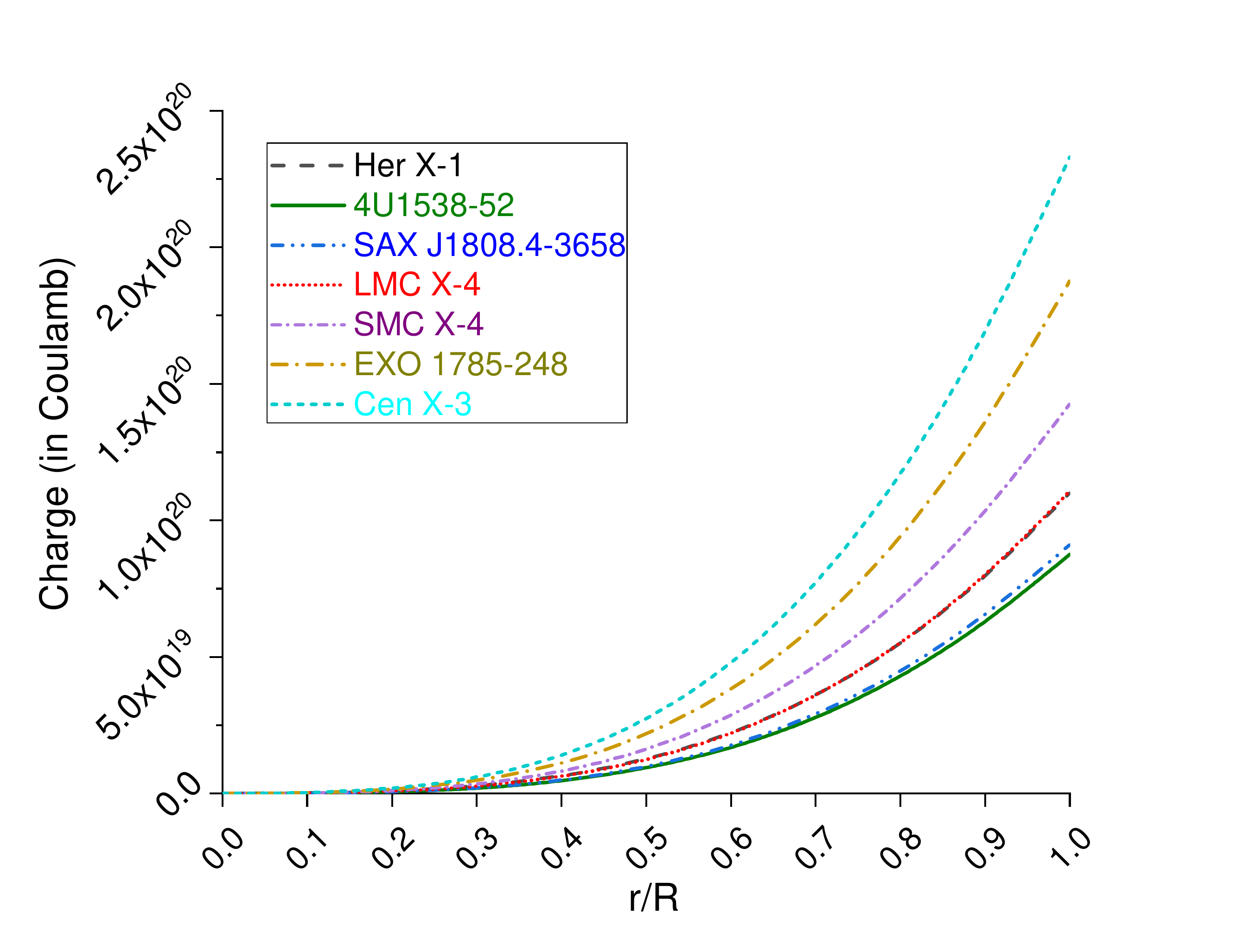}\includegraphics[width=8cm]{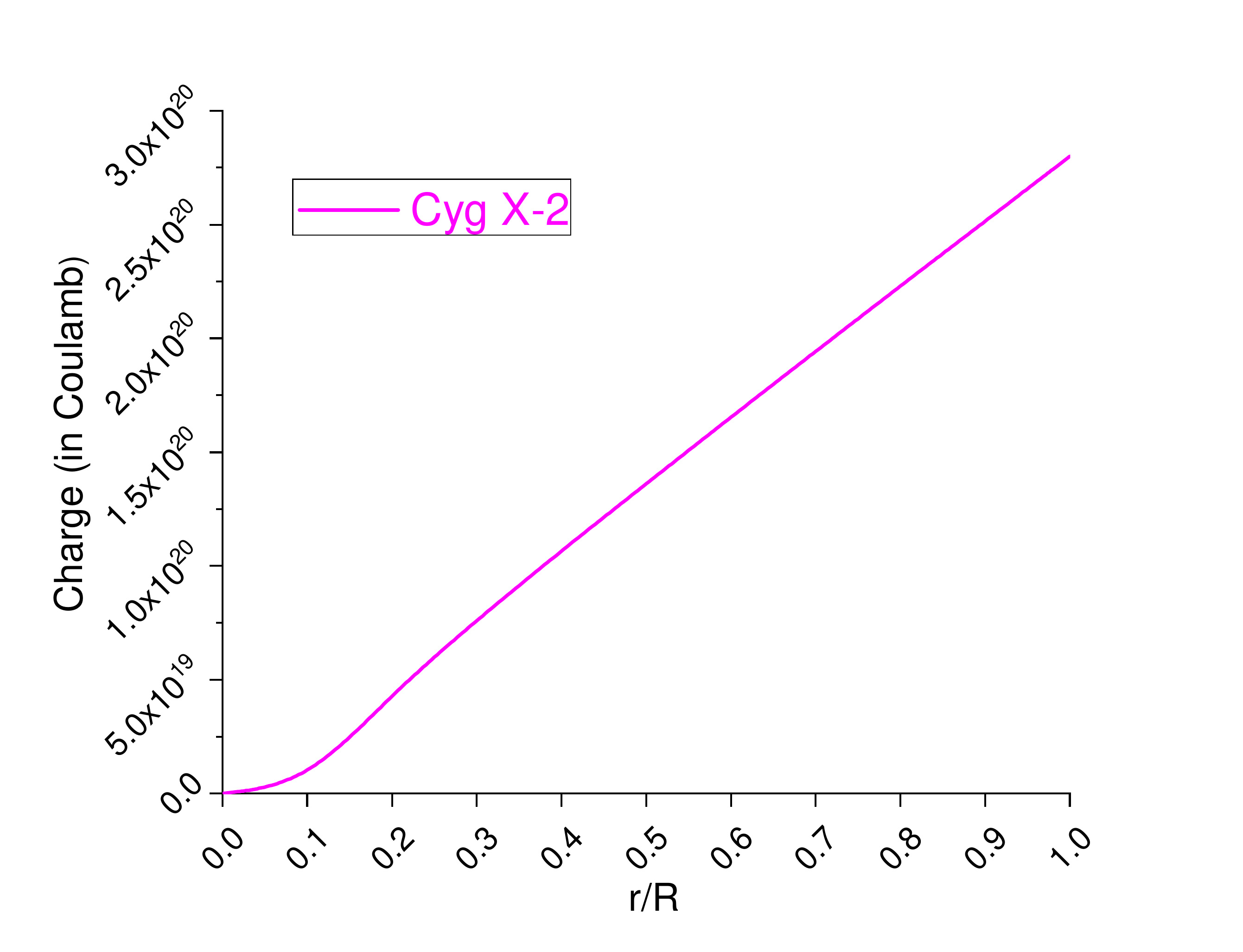}
								\caption{Variation of charge with respect to fractional radius (r/R) for star candidates Her X-1, 4U 1538-52, SAX J1808.4-3658, LMC X-4, SMC X-4,  EXO 1785-248, Cen X-3  ($K<0$) and Cyg X-2 ($K>1$).}\label{c}
						\end{figure}
					The fig (\ref{c}), clearly states that the electric field given by eq. (\ref{charge}) is positive and increasing towards the surface for each star candidate. Along with this, the charge at centre is zero and attains its maximum value at the boundary.\\
					Ray et al. \cite{ray} have demonstrated that the global balance of the forces allows a huge charge($10^{20}$C) to be available inside a compact star. 
					Referring to the Table \ref{t2}, we can say that, in this model the net charge is effective to balance the mechanism of the force.
					
					\subsection{\textbf{Charge density}}
					\begin{figure}[h]
						\includegraphics[width=8cm]{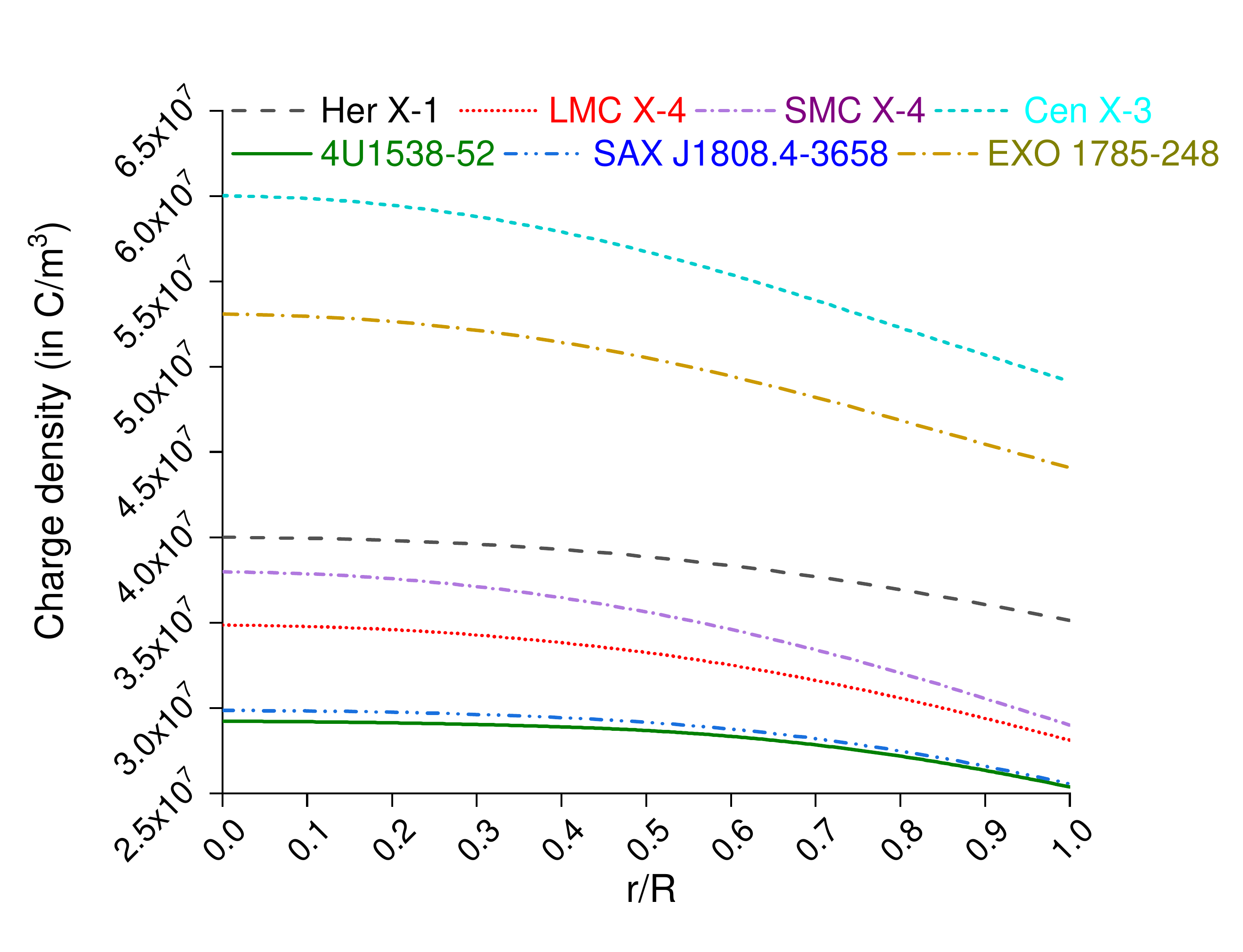}\includegraphics[width=8cm]{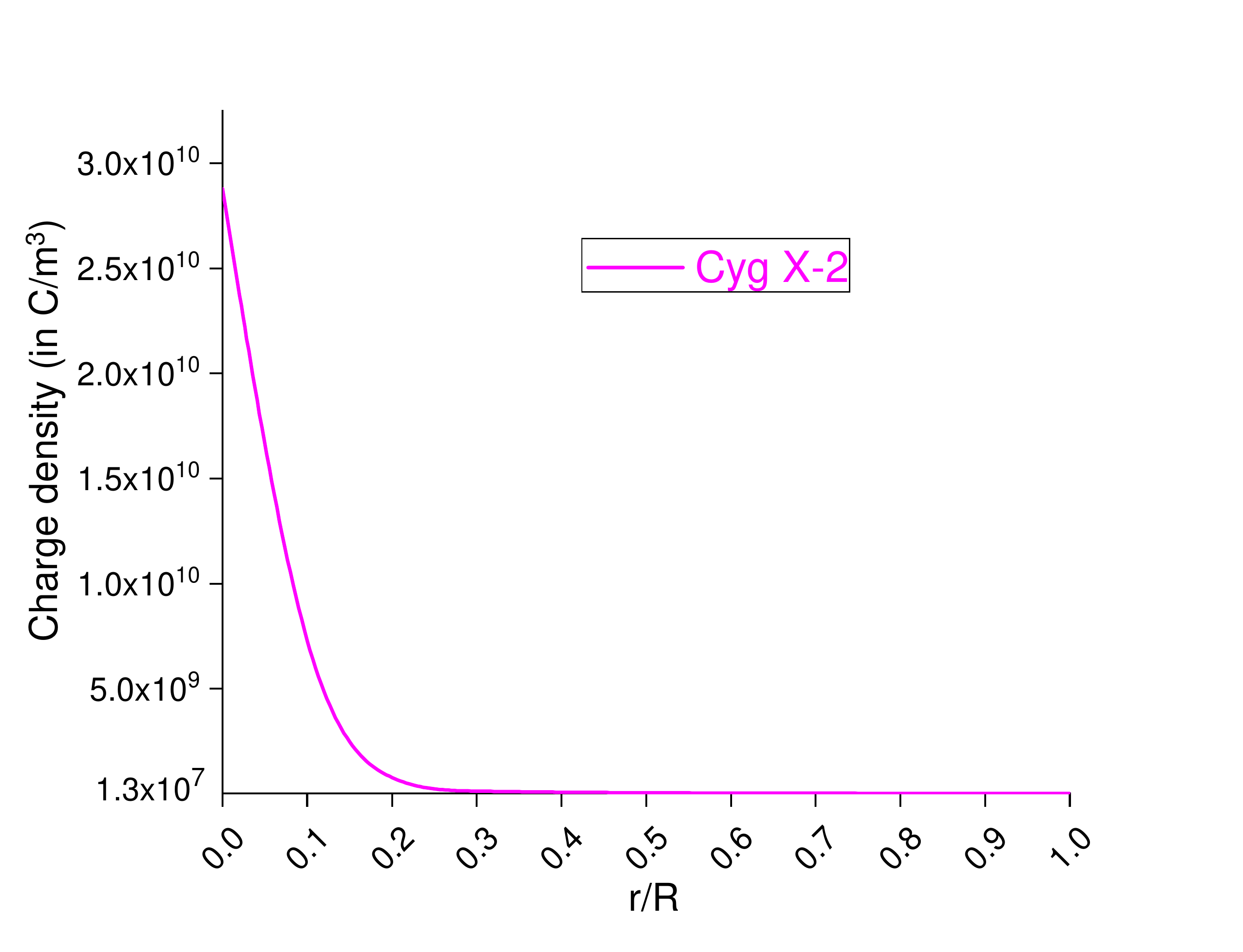}
						\caption{Variations of charge density with respect to fractional radius ($r/R$) for star candidates Her X-1, 4U 1538-52, SAX J1808.4-3658, LMC X-4, SMC X-4,  EXO 1785-248, Cen X-3  ($K<0$) and Cyg X-2 ($K>1$).} \label{sigma}
					\end{figure}
					Charge density is the amount of electric charge per unit volume. Differentiating eq. (\ref{dense}) with respect to $r$ we obtain the expression for charge density as,
					\begin{equation}
						\sigma=\frac{e^{-\lambda/2}}{4\pi r^2}\frac{dq}{dr}
					\end{equation}

					Figure \ref{sigma} shows that the proper charge density is finite at $r = 0$, regular in the interior, and evolves as a decreasing function throughout for each compact star candidates.
					\subsection{\textbf{Mass–radius relation and compactness factor}}
						By plugging Eqs. (\ref{lambda}) and (\ref{charge}) into Eq. (\ref{mass}), eventually we get
					\begin{equation}
						m(r)=\frac{(K-1)Cr^3}{2K(1+Cr^2)}+\frac{C^2r^5}{4K(1+Cr^2)^2}\Big[ \frac{5}{4}\frac{1}{(1-X^2)}-\frac{2a_1}{X^2(a_1+a_2X)}(1-X^2)\frac{7}{4}\Big].
					\end{equation}
					
\begin{figure}[h]
					\includegraphics[width=8cm]{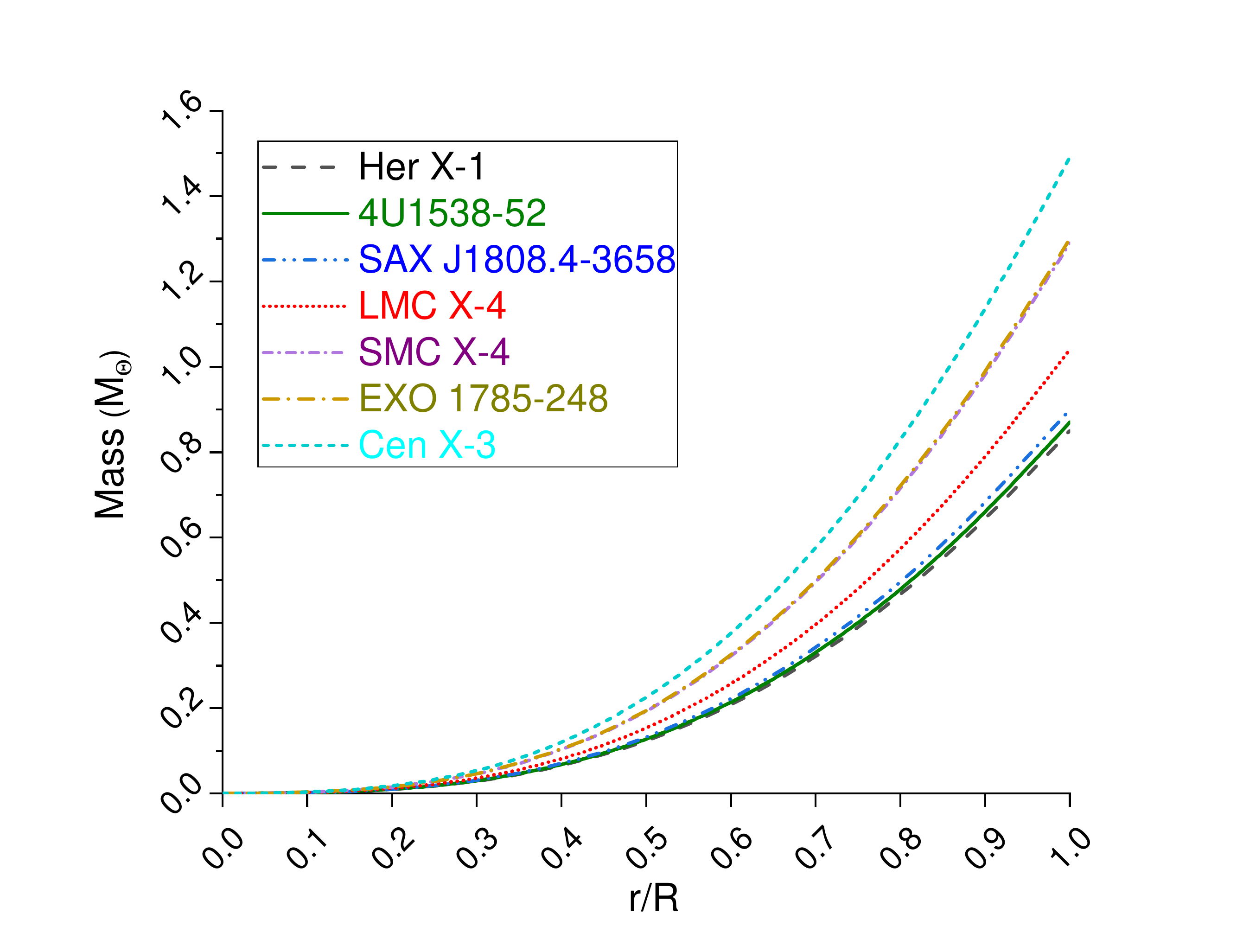}\includegraphics[width=8cm]{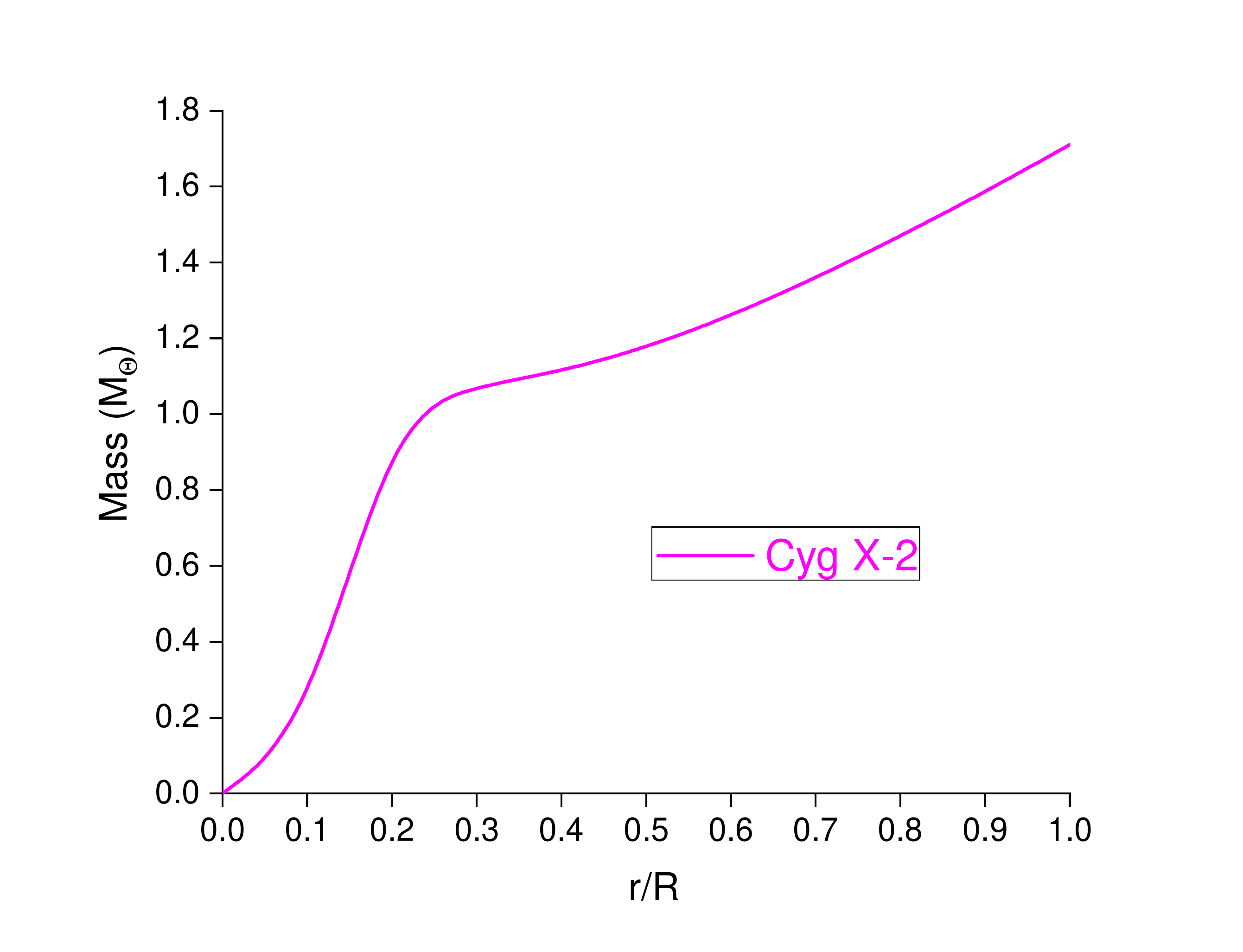}
					\caption{Variation of mass with respect to fractional radius (r/R) for star candidates Her X-1, 4U 1538-52, SAX J1808.4-3658, LMC X-4, SMC X-4,  EXO 1785-248, Cen X-3  ($K<0$) and Cyg X-2 ($K>1$).}\label{m-r}
				\end{figure}

					In Fig. \ref{m-r}, the mass function is plotted against the radius and the profile indicates an increasing function with increase of radius. For physically viable models, the ratio of the mass to that of radius of a compact star model cannot be arbitrarily large. According to Buchdahl \cite{buchdahl}, the ratio of mass to the radius for a perfect fluid compact star should satisfy the inequality $\frac{2M}{R} < \frac{8}{9}$. 
					The compactness factor $\mu(r) = \frac{m(r)}{r}$ can be computed as,
					\begin{equation}
						\mu(r)=\frac{(K-1)Cr^2}{2K(1+Cr^2)}+\frac{C^2r^4}{4K(1+Cr^2)^2}\Big[ \frac{5}{4}\frac{1}{(1-X^2)}-\frac{2a_1}{X^2(a_1+a_2X)}(1-X^2)\frac{7}{4}\Big].
					\end{equation}
				\begin{figure}[h]
					\includegraphics[width=8cm]{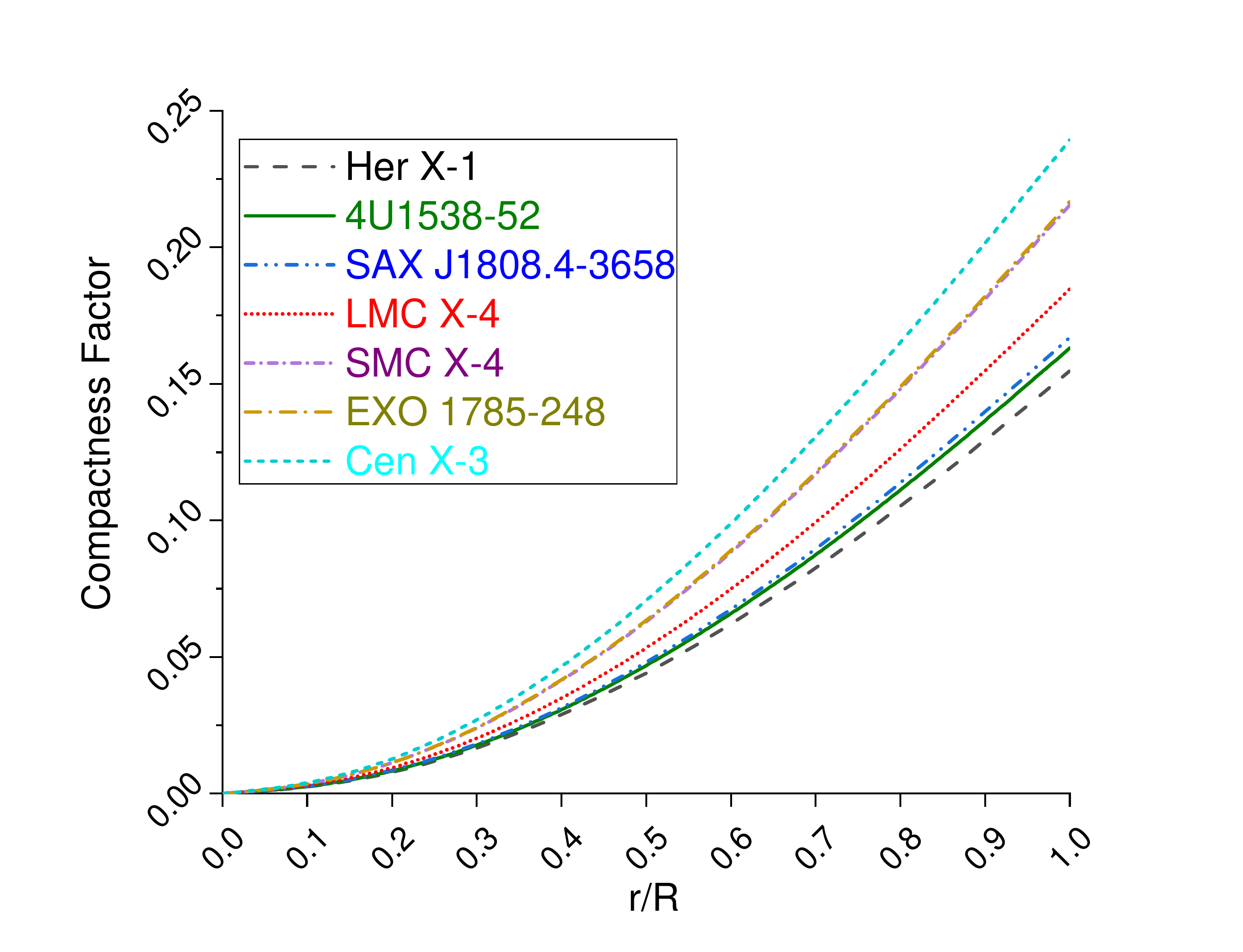}\includegraphics[width=8cm]{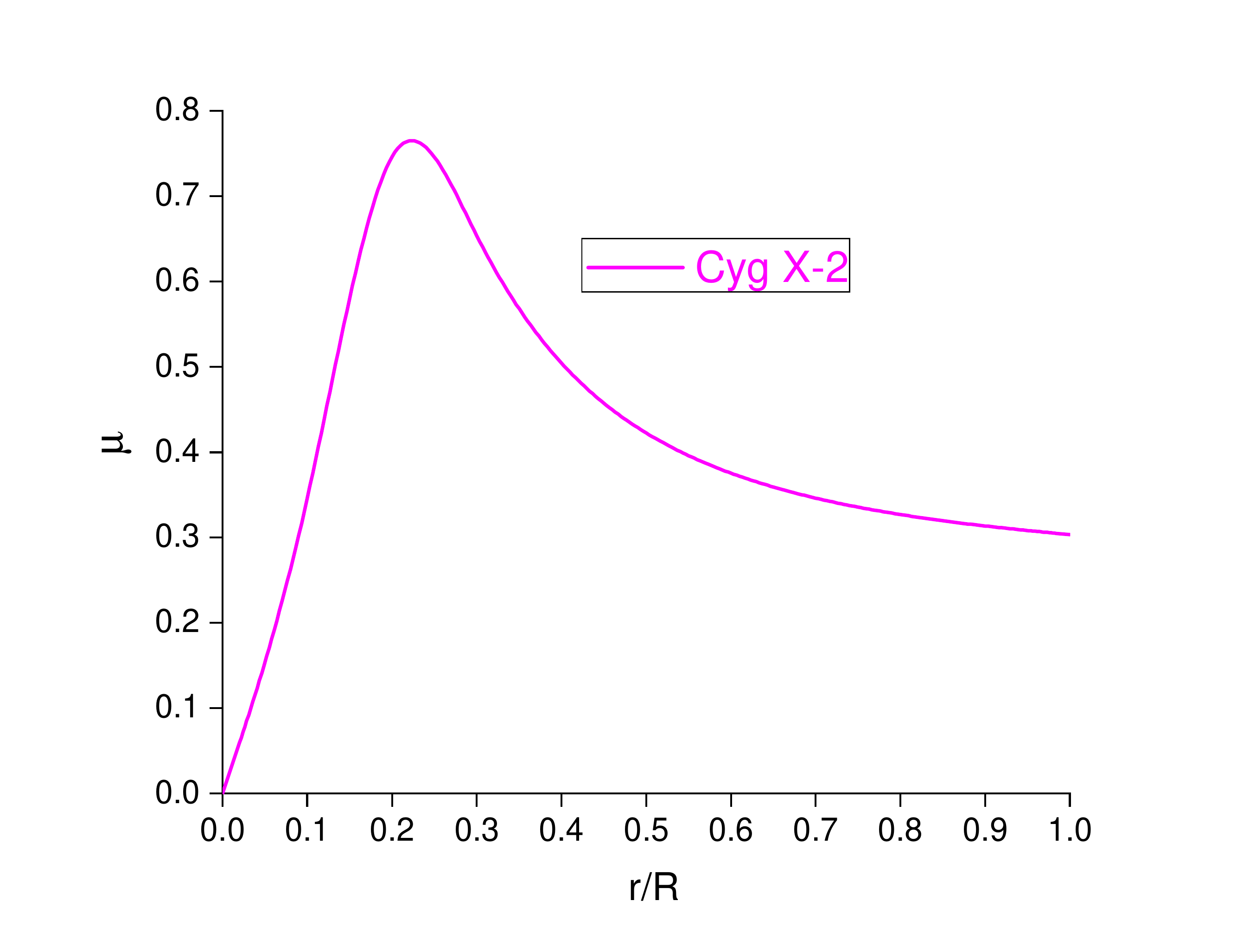}
					\caption{Compactness factor with respect to fractional radius (r/R) for star candidates Her X-1, 4U 1538-52, SAX J1808.4-3658, LMC X-4, SMC X-4,  EXO 1785-248, Cen X-3  ($K<0$) and Cyg X-2 ($K>1$).}\label{compact}
				\end{figure}
					The compactness factor is plotted in Fig. \ref{compact}. We can see in Table \ref{t1} that for each compact star candidate we have considered here, the value of $\mu$ is consistent with the condition of Buchdahl.
					However, Böhmer and Harko \cite{bohmer} have given the generalized expression of lower bound for a charged compact object as follows:
					\begin{equation}
						\frac{Q^4+18R^2Q^2}{12R^4+R^2Q^2}\le \frac{2M}{R}
					\end{equation}
					Subsequently, Andreasson \cite{andreasson} showed that, for a charged sphere, the model must satisfy the following inequality
					\begin{equation}
						\sqrt{M} \le \frac{\sqrt{R}}{3}+\sqrt{\frac{R^2+3Q^2}{9R}}
					\end{equation}
					We, therefore, conclude from the above two conditions that $\frac{2M}{R}$ must satisfy the following inequality:
					\begin{equation}
						\frac{Q^2}{R^2}\left(\frac{Q^2+18R^2}{12R^2+Q^2}\right)\le \frac{2M}{R} \le \frac{2}{R}\left[ \frac{\sqrt{R}}{3}+\sqrt{\frac{R^2+3Q^2}{9R}}\right]^2.
						\label{boundM}
					\end{equation}
					\begin{table}[h]
						\centering
						\caption{Upper and lower bound of $\mu$ for the compact star candidates.}
						\label{bound}
						 \begin{tabular}{|l|cccccccc|}
							\hline
							Compact Star	&Her X-1&4U 1538-52 &SAX J1808.4-3658&LMC X-4&SMC X-4 &EXO 1785-248&Cen X-3&Cyg X-2 \\ 
							\hline
							Lower bound &0.010176926&	0.006827579&	0.007248024&	0.00982036&	0.014368244&	0.02480304&	0.035576893&	0.062467518\\				
							Upper  bound &	0.453448858&	0.450494487&	0.450865833&	0.453134756&	0.457133708&	0.466250717&	0.475582871&	0.498546049 \\
							\hline
						\end{tabular}
					\end{table}
				Using eq. (\ref{boundM}), we have obtained the ranges for compactness factor in Table \ref{bound}. We can observe from Table \ref{t1} and Table \ref{bound} that the value of $\mu$, for the considered compact star candidates, lie in this range. Thus, $\mu$ for each compact star candidate is consistent with the condition (\ref{boundM}) for a stable configuration.
					%The effective gravitational mass is given by,
					%	\begin{equation}
					%	M_{eff}=\frac{1}{2}\int_{0}^{R}\left(c^2\kappa\rho+\frac{q^2}{r^4}\right)dr=\frac{R}{2}\left(1-e^{-\lambda(R)}\right),
					%	\label{meff}	
					%\end{equation}
					%	which for our model can be  expressed as
					% \begin{equation}
					% M_{eff}=\frac{(K-1)CR^3}{2K(1+CR^2)}.
					%\end{equation}
				
					\subsection{\textbf{Gravitational redshift and Surface redshift}}
				Let's consider the gravitational redshift $z_g$ of compact objects with help of the definition, $z_g=\frac{\lambda_0-\lambda_e}{\lambda_e}$, where  $\lambda_0$ is the observed wavelength and $\lambda_e$ is the emitted wavelength at the surface of a non-rotating star. Thus, the gravitational redshift from the surface of the star, as measured by a distant observer, is given by	
				\begin{equation}
					z=\frac{1}{\sqrt{|e^{\nu(R)}|}}-1 = \Big(1-\frac{2M}{R}+\frac{Q^2}{R^2}\Big)^{-1/2}-1
				\end{equation}
				
				Gravitational redshift is a phenomenon in which electromagnetic waves or photons seem to lose energy when it climbs out of a gravitational well. The surface redshift depends on the surface gravity, i.e., on the overall mass and radius of stellar object. The gravitational (interior) redshift $z_g(r)$ and surface redshift $z_s(r)$ are defined as, 
				\begin{eqnarray}
					z_g(r)&=&\frac{1}{\sqrt{|e^{\nu(r)}|}}-1
					\label{zg}\\
					z_s(r)&=&\frac{1}{\sqrt{1-2\mu+\frac{q^2}{r^2}}}-1
					\label{zs}	
				\end{eqnarray}
				
				If a photon comes out from center to surface, it has to travel a denser region and longer path, which leads to more dispersion and a great loss of energy. Whereas, when a photon comes out from near the surface, it has to travel a comparitively less denser region and shorter path, therefore, it goes through less dispersion and less energy loss takes place. Hence, the interior redshift is minimum at the surface and maximum at the center.
				
				As radius slightly increases with increase in mass resulting into more surface gravity, the surface redshift is maximum at the surface and decreases towards the center. Moreover, at the surface of stars, $z_s(R)=z_g(R)=z$, implying that minimum value of inteior redshift is the maximum for surface redshift.
				
\begin{figure}[h]
						\includegraphics[width=8cm]{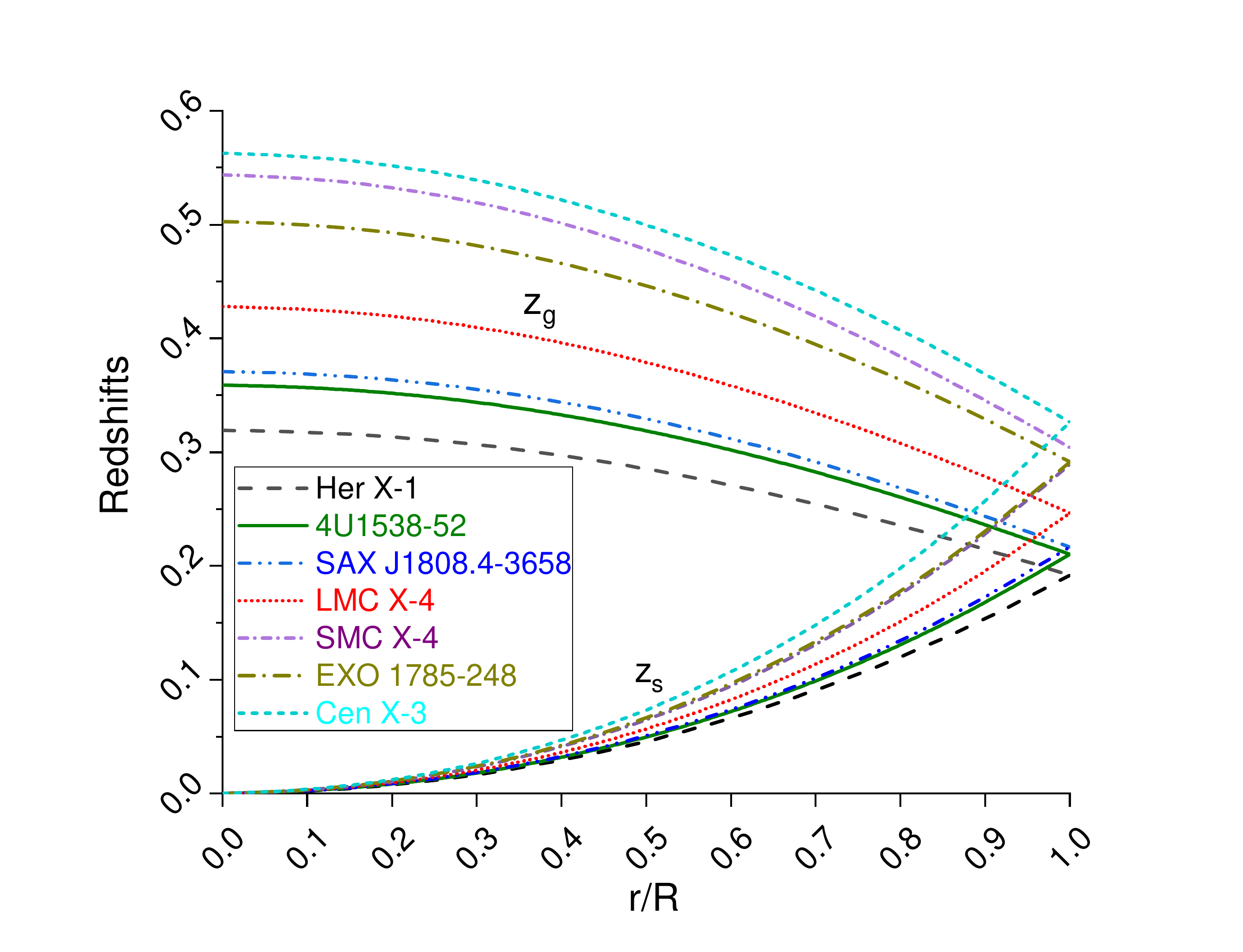}\includegraphics[width=8cm]{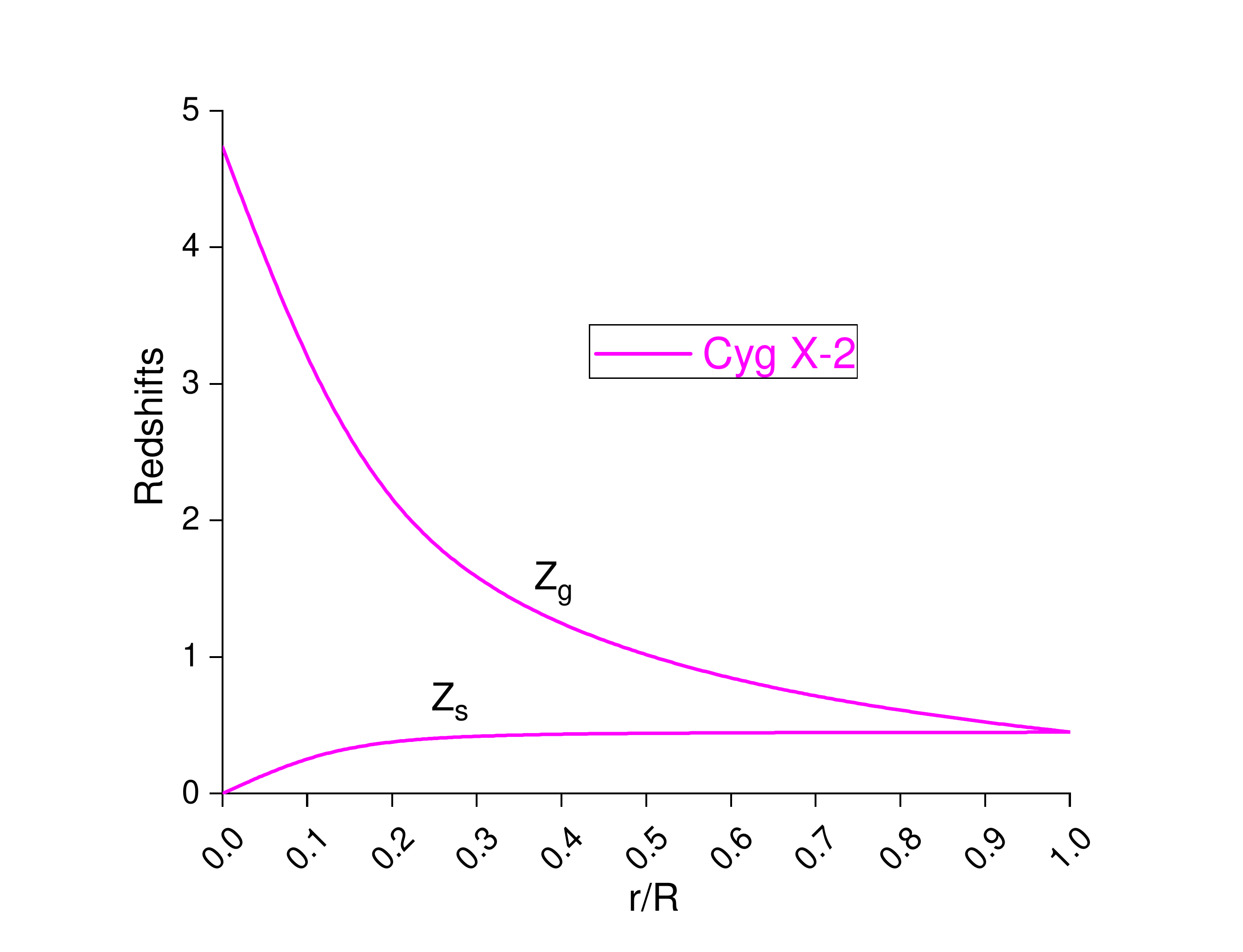}
						\caption{Variation of redshift with respect to fractional radius (r/R) for star candidates Her X-1, 4U 1538-52, SAX J1808.4-3658, LMC X-4, SMC X-4,  EXO 1785-248, Cen X-3  ($K1$).}\label{red}
				\end{figure}

				To explore the behaviour of the redshifts, we have provided its graphical representation in Fig. \ref{red}.  We can see through figure that the redshifts has no sigularity throughout its configuration.
				
				For an isotropic star a constraint on the gravitational redshift for perfect fluid spheres is given by $z_s< 2$  \cite{buchdahl, straumann}. As can be seen in Table \ref{t2}, for the constants mentioned in Table \ref{t1}, surface redshift of the star candidates have values less than 2. 
					
					\subsection{\textbf{Causality Condition}}
					Now, we are going to analyse the speed of sound propagation $v_s^2$, which is given by
					\begin{equation}
						{v_s}^2=\frac{dp}{d\rho}=\frac{\frac{X^2}{K(X^2-1)} \frac{D}{P_2  P_5}+\frac{2}{K(1-K)(X^2-1)^2}\Big(\frac{P_1  P_2+P_3  P_4}{P_2  P_5}-1\Big)+D_7+D_8}{D_6-D_7-D_8}
					\end{equation}
					\begin{figure}[H]
					\includegraphics[width=8cm]{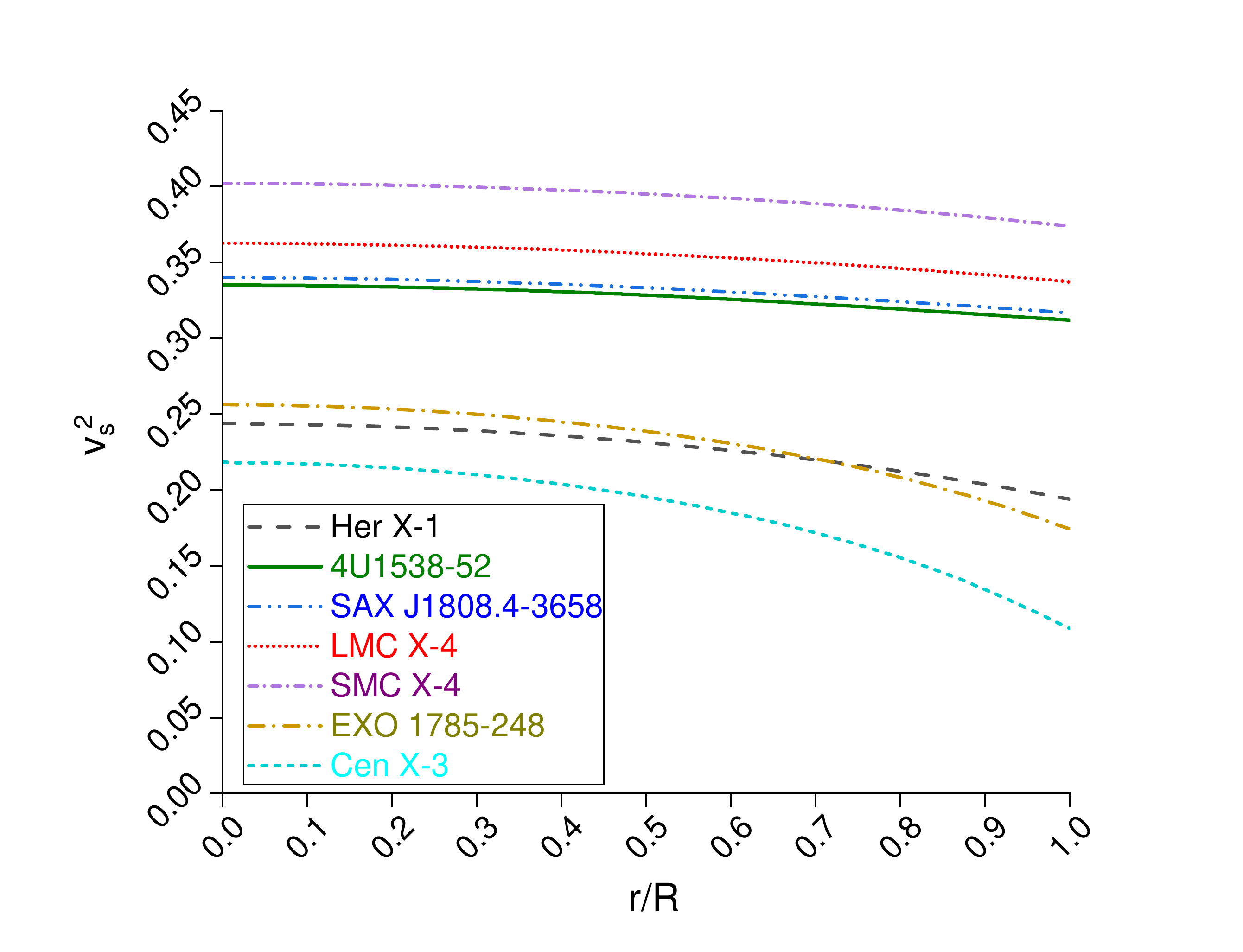}\includegraphics[width=8cm]{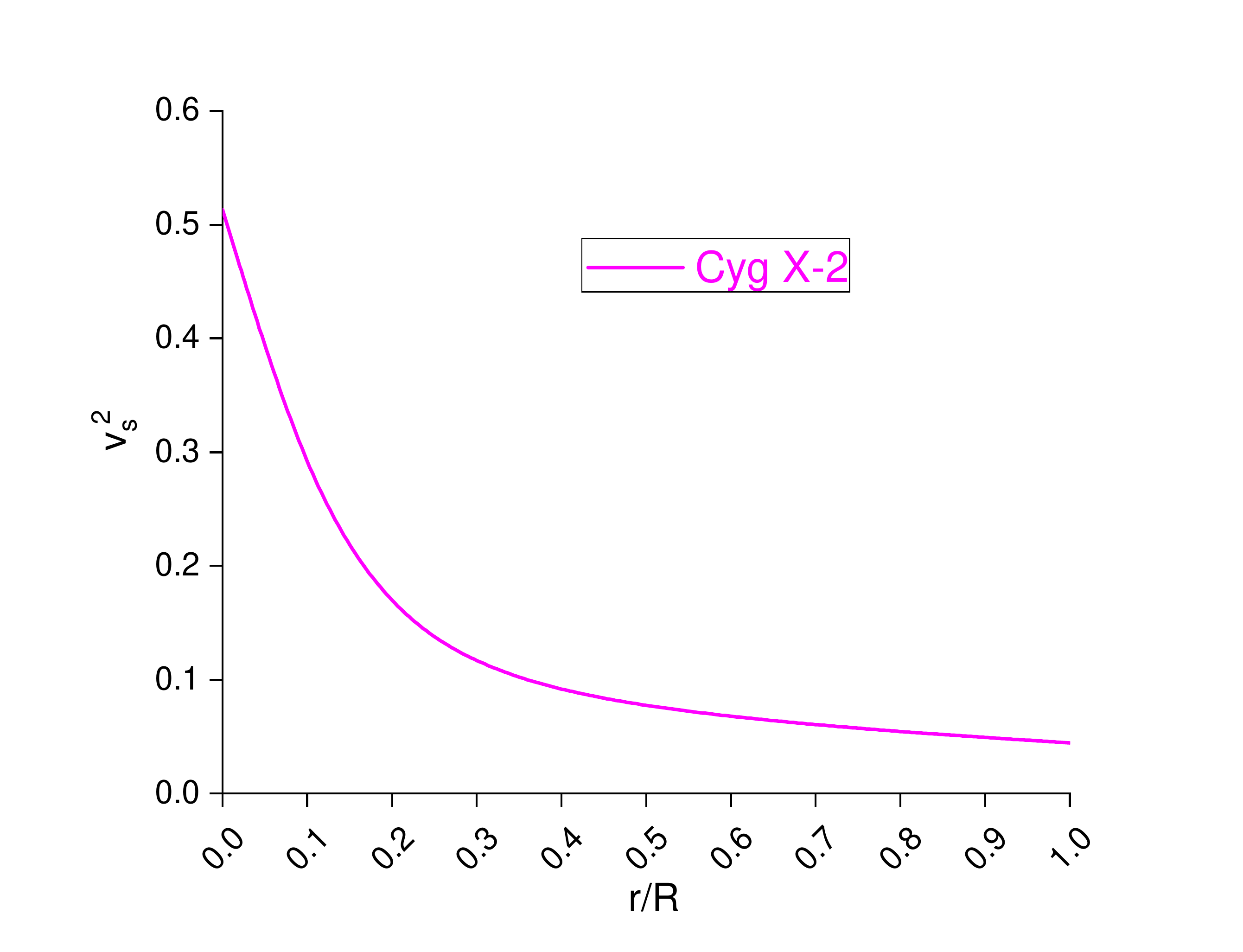}
					\caption{Variation of velocity of sound of compact star candidates Her X-1, 4U 1538-52, SAX J1808.4-3658, LMC X-4, SMC X-4,  EXO 1785-248, Cen X-3  ($K<0$) and Cyg X-2 ($K>1$)  with respect to fractional radius (r/R).}\label{v}
				\end{figure}
					Naturally the velocity of sound does not exceed the velocity of light. Thus, the sound speed must have value less than 1, as we have taken $c=1$. For a physically acceptable isotropic fluid distribution, the causality condition, i.e., $0\le {v_s}^2\le1$, must be satisfied  to achieve a stable equilibrium. It was stated by Canuto \cite{caunto} that for an ultra-high distribution of matter, the speed of sound should decrease monotonically towards the surface of the star.\\
					We have shown in Fig. (\ref{v}) that, for our charged isotropic model, the speed of sound remains less than the speed of light and for each star candidate, it decreases with increase in $r$.
				
					\subsection{\textbf{Equation of state}}
\begin{figure}[H]						
							\includegraphics[width=8cm]{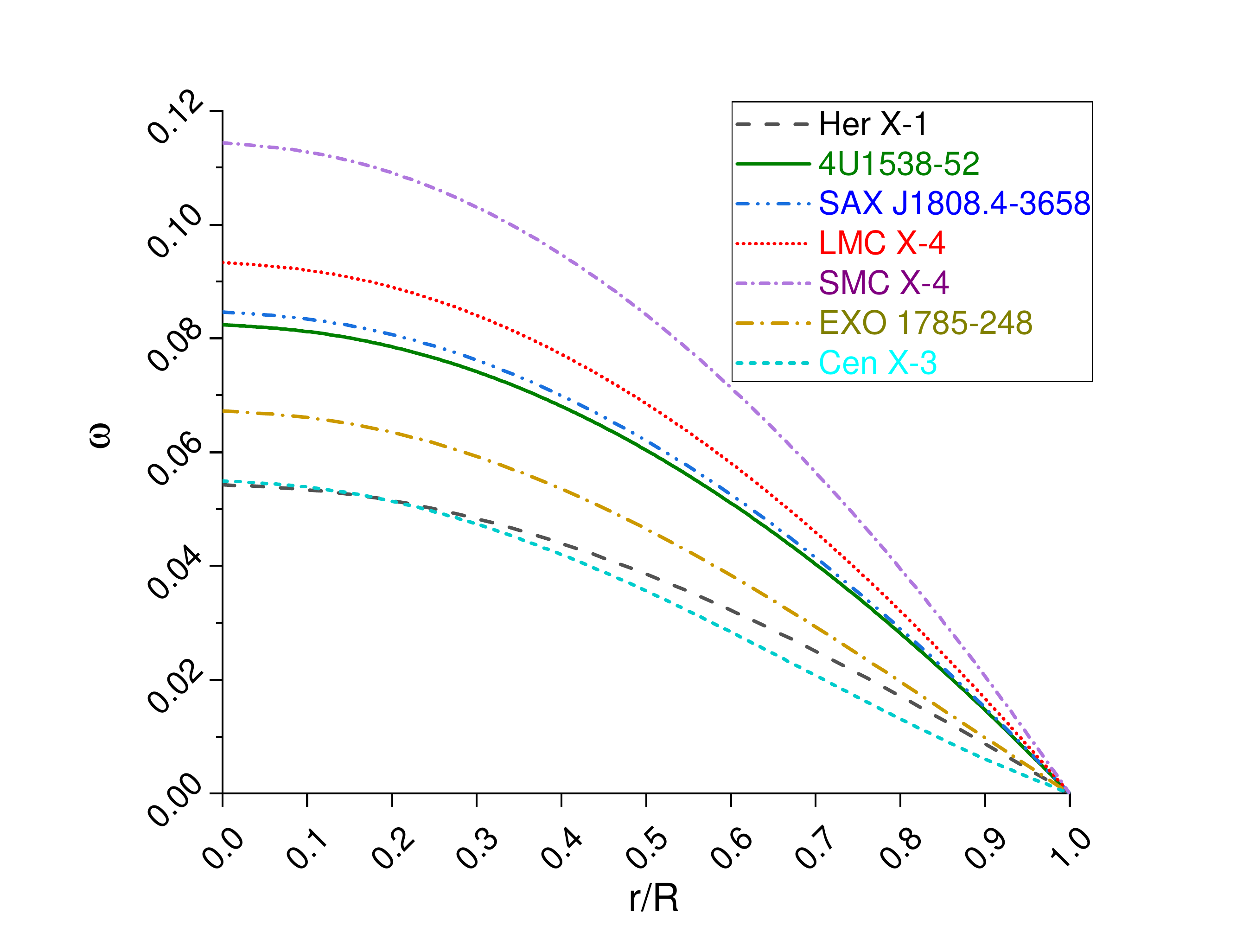}\includegraphics[width=8cm]{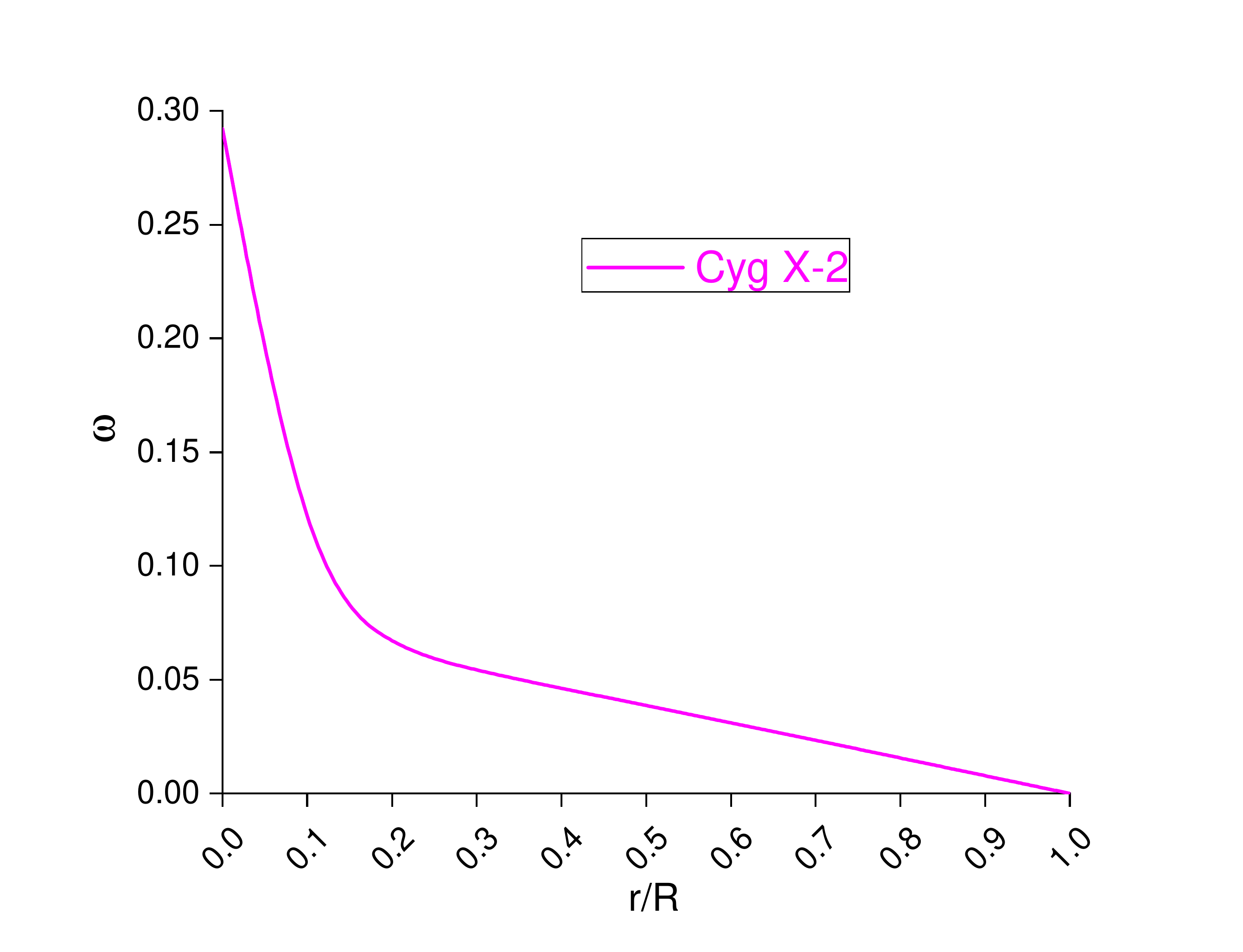}
							\caption{Ratio of pressure to that of density with respect to fractional radius (r/R) for star candidates Her X-1, 4U 1538-52, SAX J1808.4-3658, LMC X-4, SMC X-4,  EXO 1785-248, Cen X-3  ($K<0$) and Cyg X-2 ($K>1$).}\label{ratio}
					\end{figure}

					The term "Equation of state (EoS)" means a function $p(\rho)$, which estabilish a relation between the pressure $p$ and energy density $\rho$.
					Lets consider that the pressure of the charged fluid sphere is related with their energy density, by a parameter $\omega$ via the EoS, $p=\omega \rho$, which
					is given by
					\begin{equation}
						\omega=\frac{\frac{X^2}{X^2-1} \Big[\frac{P_1P_2+P_3P_4}{P_2P_5}\Big]-\frac{1}{(X^2-1)}+\frac{Cr^2}{2(1+Cr^2)^2}\Big[\frac{5}{4(1-X^2)}-\frac{2a_1(1-X^2)}{X^2(a_1+a_2X)}+K-\frac{7}{4}\Big]}{\frac{(K-1)(3+Cr^2)}{(1+Cr^2)^2}-\frac{Cr^2}{2(1+Cr^2)^2}\Big[\frac{5}{4(1-X^2)}-\frac{2a_1(1-X^2)}{X^2(a_1+a_2X)}+K-\frac{7}{4}\Big]}
					\end{equation}
			In Fig. (\ref{ratio}), the factor $\omega$ has been plotted against the fractional radial coordinate (r/R). We can see in this figure that, throughout
			the interior of stars, the ratio $\omega = p/rho$ is less than unity. This result implies that, inside the stars, densities are dominating over the corresponding pressures everywhere  and therefore the underlying fluid distribution is non-exotic in its nature \cite{rahman}.

%	Also, in the figure it seems that we can form separate EoS for stars from our model. Although, the EoS for elementary matter should be the same for all of the stars, it is possible that two compact stars have quite different EoS \cite{li}. The EoS is different for different compact star candidates shows that the internal constituent matters of the stars are in different proportions. Thus, we can say that the general EoS for stars in the model must be the same but they can take separate forms for different stars. However, we can note that effectively only three different EoSs (two for $K<0$ and one for $K>1$) occurs in this model. EoS of LMC X-4 is quite similar to EXO 1785-248 and remaining stars for $K<0$ has similar EoSs.

					\subsection{\textbf{Energy Conditions}}
					It is justifiable to expect this model to satisfy the energy conditions within the framework of general relativity. There exists a linear relationship between energy density and pressure, obeying certain restrictions. This is termed as energy conditions. To enhance our investigation on the structure of  relativistic space-time,	let's examine the following conditions \cite{EngCond}:
					\begin{enumerate}
						\item Dominant energy condition  (DEC): $\rho-p\ge 0$
						\item Null energy condition (NEC): $\rho+\frac{q^2}{8\pi r^4}\ge0$ 
						\item Weak energy condition (WEC): $\rho-p+\frac{q^2}{4\pi r^4}\ge0$
						\item Strong energy condition (SEC): $\rho-3p+\frac{q^2}{4\pi r^4}\ge0$
					\end{enumerate}
					\begin{figure}[H]						
							\includegraphics[width=8cm]{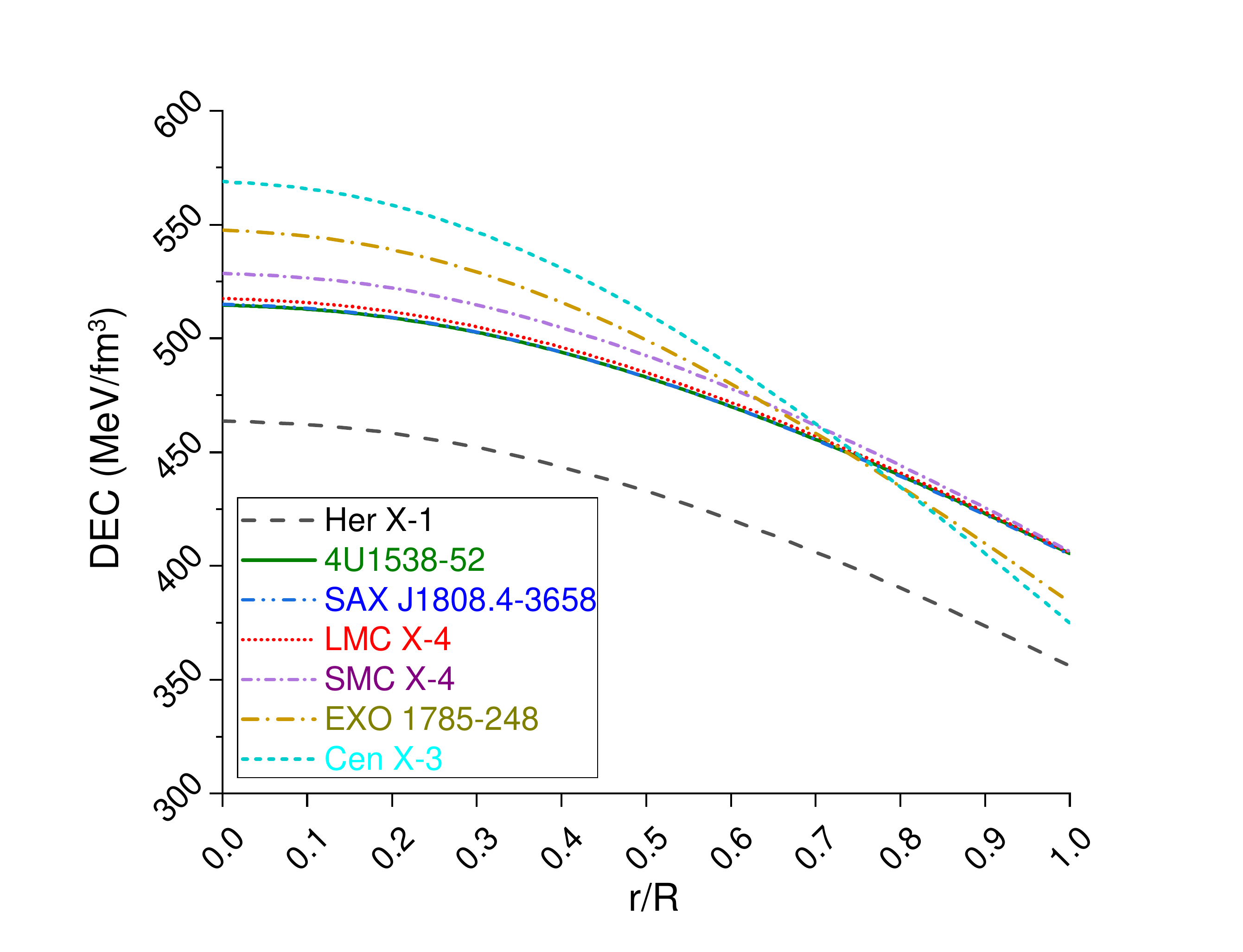}\includegraphics[width=8cm]{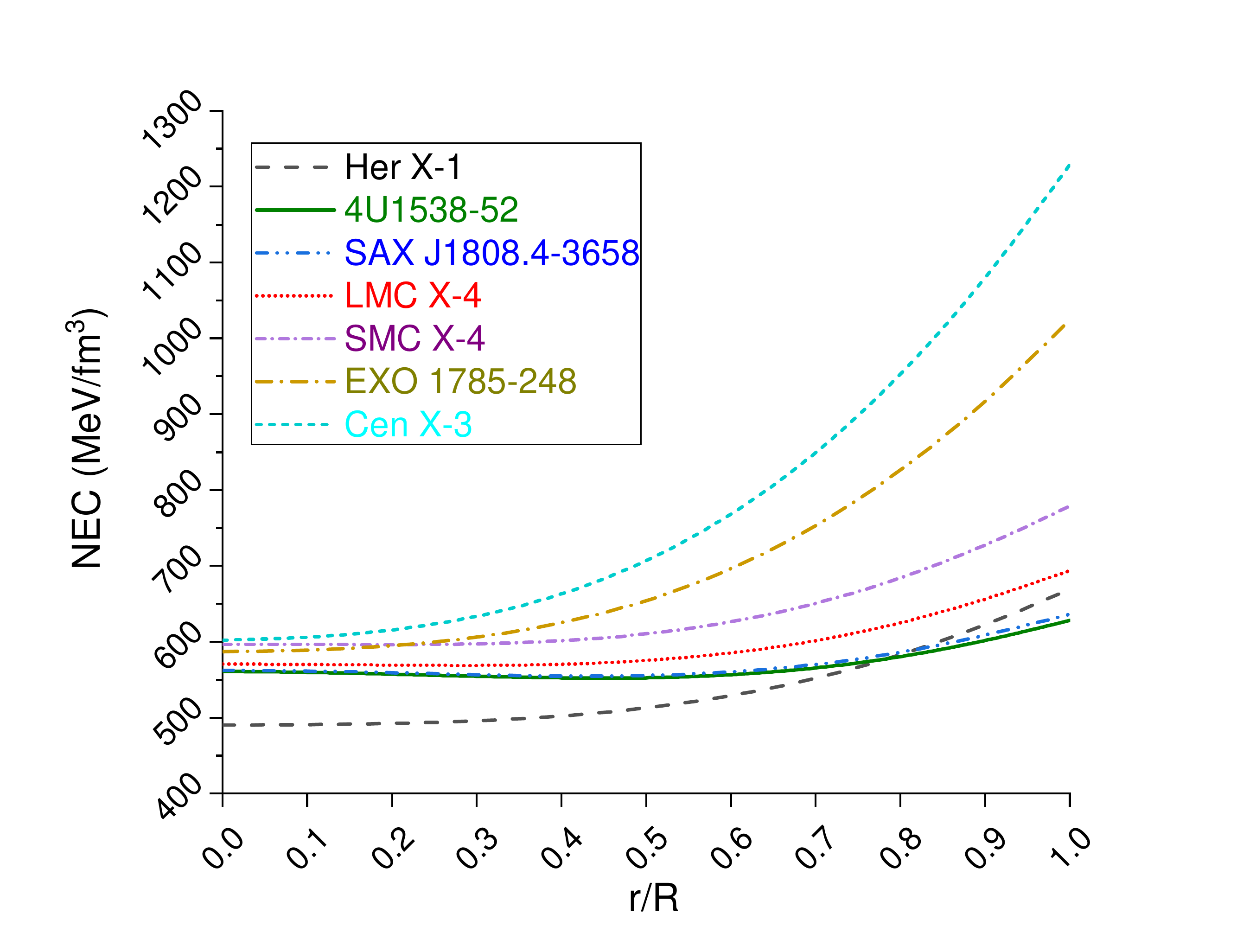}
							\includegraphics[width=8cm]{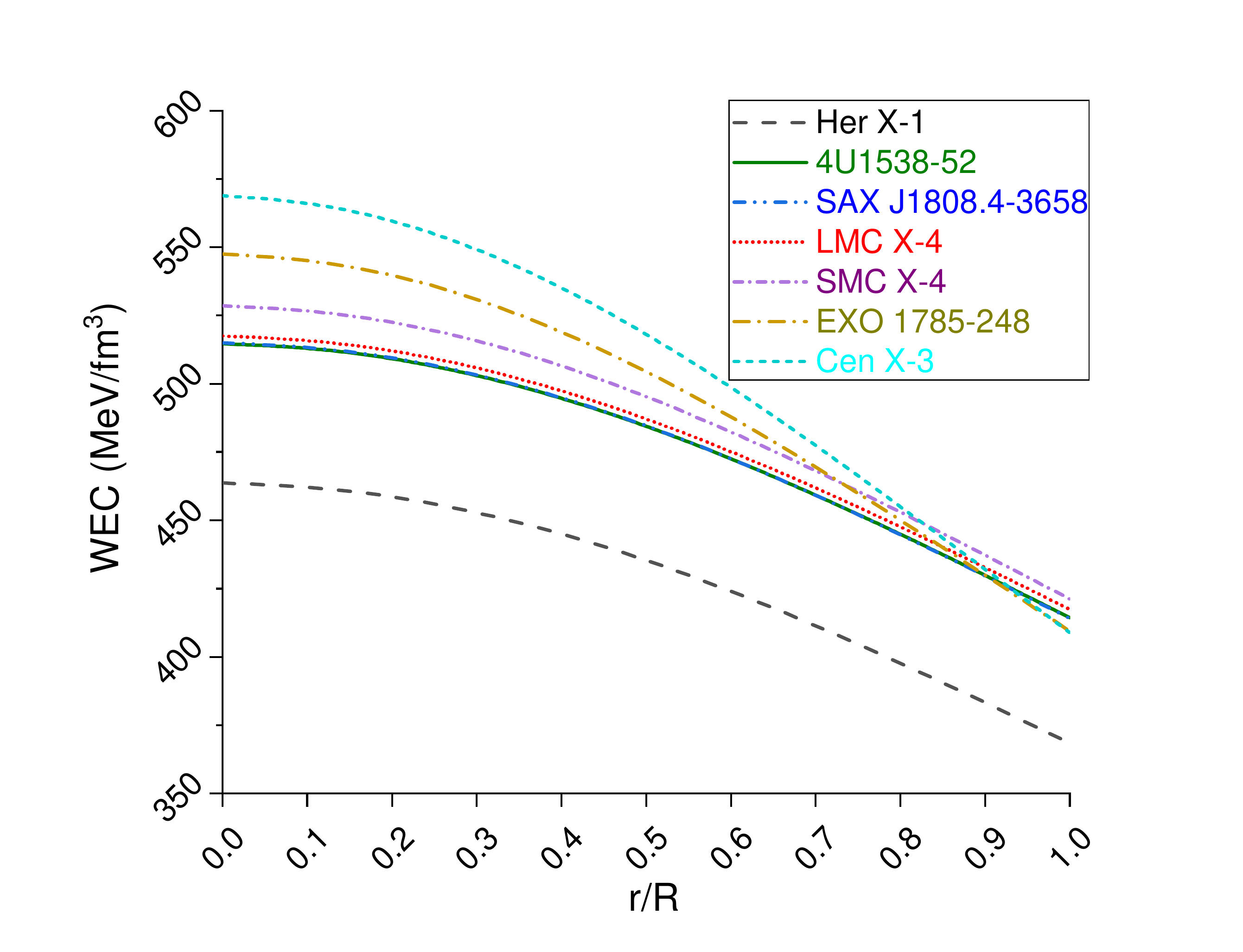}\includegraphics[width=8cm]{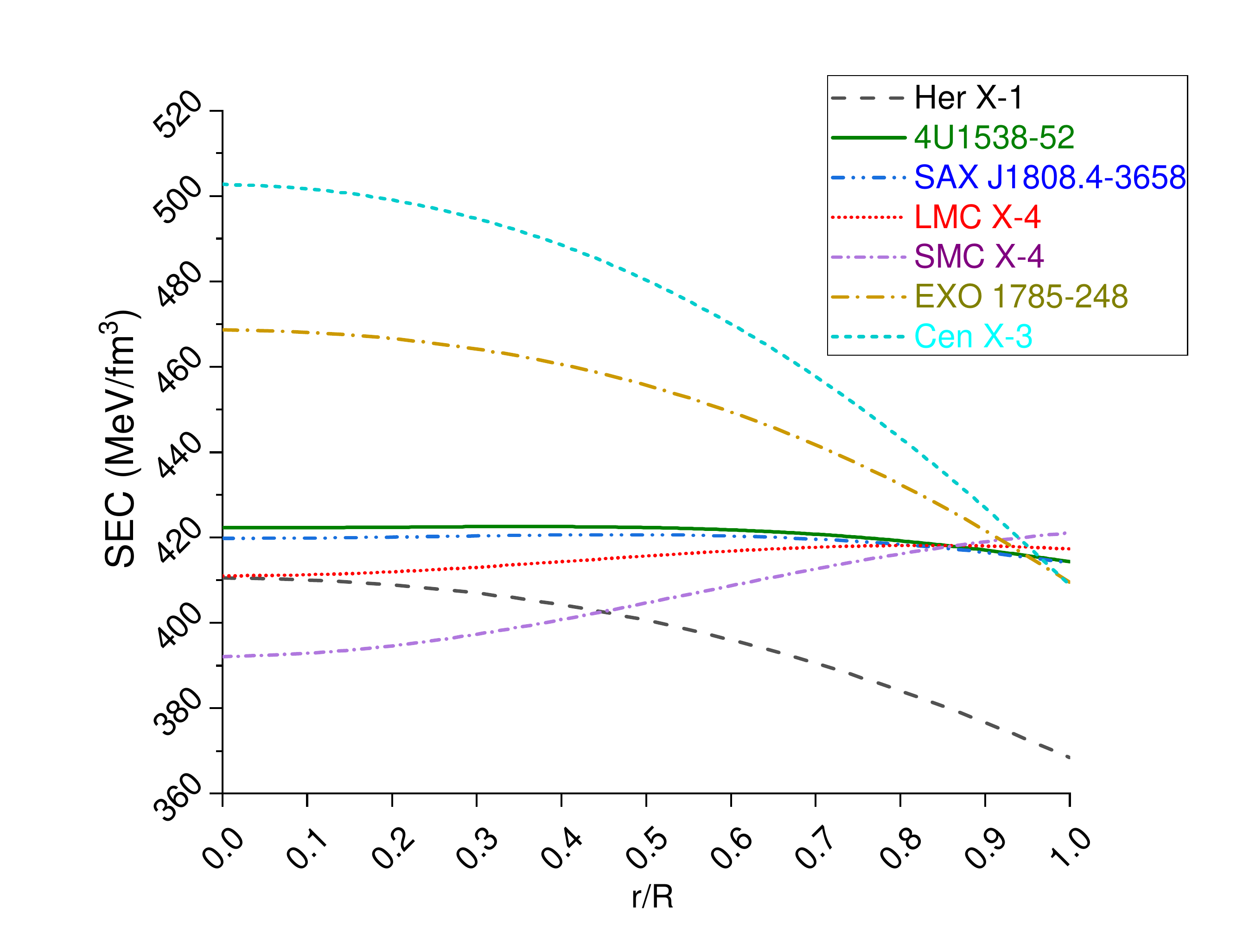}
							\begin{center}
								\includegraphics[width=8cm]{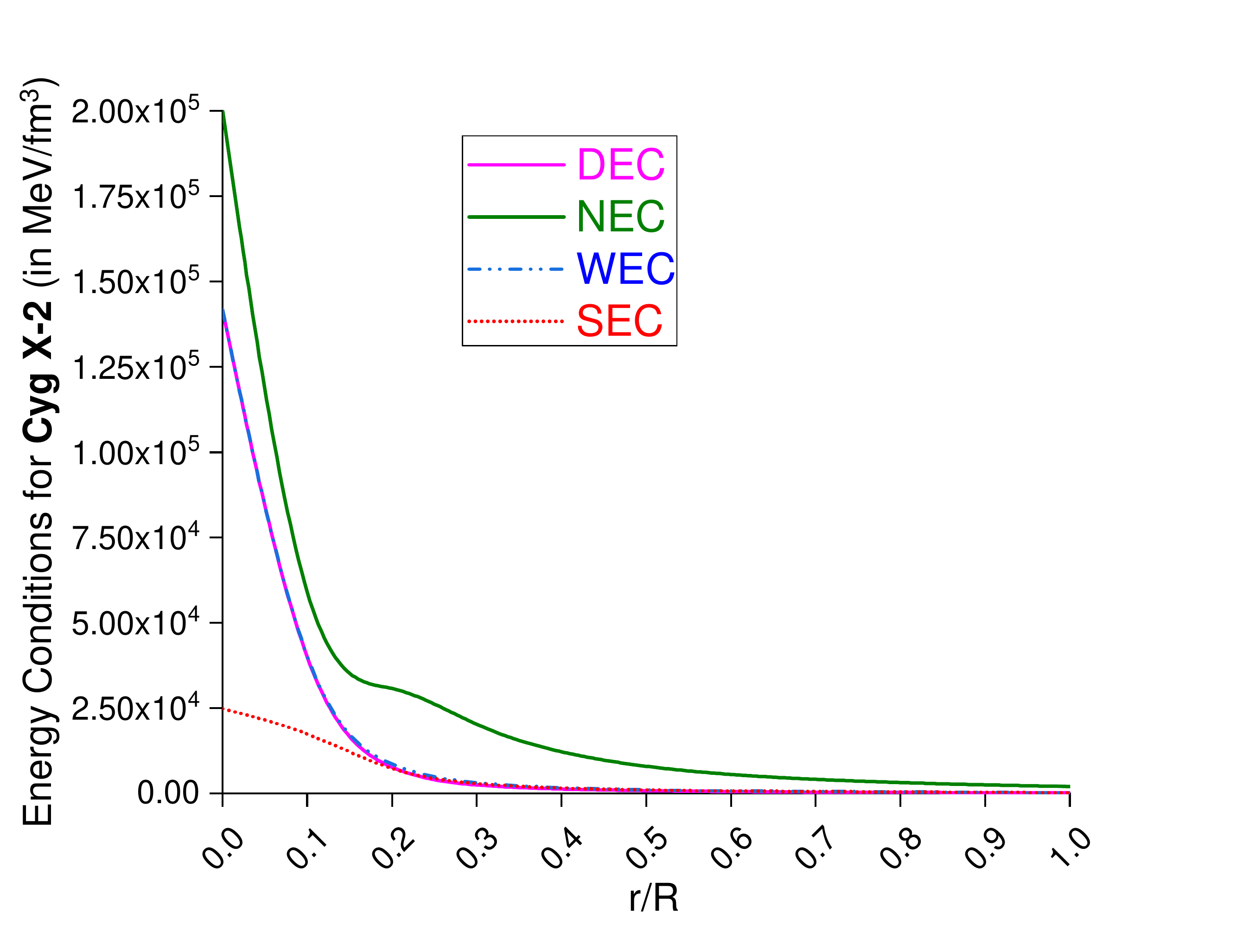}
							\end{center}
							\caption{Energy conditions on the system with respect to fractional radius ($r/R$) for star candidates Her X-1, 4U 1538-52, SAX J1808.4-3658, LMC X-4, SMC X-4,  EXO 1785-248, Cen X-3  ($K<0$) and Cyg X-2 ($K>1$).} \label{ec}
					\end{figure}
					Nature of energy conditions for considered star candidates has been shown in Fig. (\ref{ec}). We can see that all the energy conditions are satisfied throughout the interior region of the spherical distribution.
					
					\subsection{\textbf{Generalized Tolman-Oppenheimer-Volkov Equation}}
					A star remains in hydrostatic equilibrium under different forces, namely, gravitational force ($F_{g}$), hydrostatic force ($F_{h}$) and electric force ($F_{e}$). Let's consider the generalized Tolman-Oppenheimer-Volkoff equation in the presence of  charge \cite{tov3}
					\begin{equation}
						\frac{-M_{G}(\rho+p)}{r^2}e^{(\lambda-\nu)/2}-\frac{dp}{dr}+\sigma \frac{q}{r^2}e^{\lambda/2}=0 ,
						\label{tov}
					\end{equation}
					where $M_{G}(r)$ is  the effective gravitational mass of the star within radius $r$ and is defined by 
					\begin{equation}
						M_{G}(r)=\frac{1}{2}r^2 \nu ' e^{(\nu-\lambda)/2}
					\end{equation}
					Substituting the value of $M_{G}(r)$ in eq. (\ref{tov}), we obtain,
					\begin{equation}
						F_{g} + F_{h}+F_{e}=0
					\end{equation}
			where,
					\begin{eqnarray}
					\nonumber	F_{g}&=&-\frac{\nu '}{2}(\rho+p) = -\frac{C^2r}{16 \pi} \Big[\frac{P_1  P_2+P_3  P_4}{P_2  P_5}\Big] \Big[ \frac{2}{K(K-1)(X^2-1)^2}+ \frac{X^2}{K(X^2-1)} \frac{P_1  P_2+P_3  P_4}{P_2  P_5}\Big],\\
					\nonumber	F_{h}&=&-\frac{dp}{dr}= -\frac{C^2r}{8\pi}\Big[\frac{X^2}{K(X^2-1)^2} \frac{D}{P_2  P_5}+\frac{2}{K(1-K)(X^2-1)^2}\Big(\frac{P_1  P_2+P_3  P_4}{P_2  P_5}-1\Big)+D_7+D_8 \Big],\\
					\nonumber	F_{e}&=&\sigma \frac{q}{r^2}e^{\lambda/2}= \frac{1}{8\pi r^4}\frac{dq^2}{dr}= \frac{C^2r}{8\pi}\Big[\frac{3+Cr^2}{K(1+Cr^2)^3}\Big\{ \frac{5}{4(1-X^2)}-\frac{2a_1(1-X^2)}{X^2(a_1+a_2X)}+K-\frac{7}{4}\Big\}+D_8 \Big].
					\end{eqnarray}
				
\begin{figure}[H]
						\includegraphics[width=8cm]{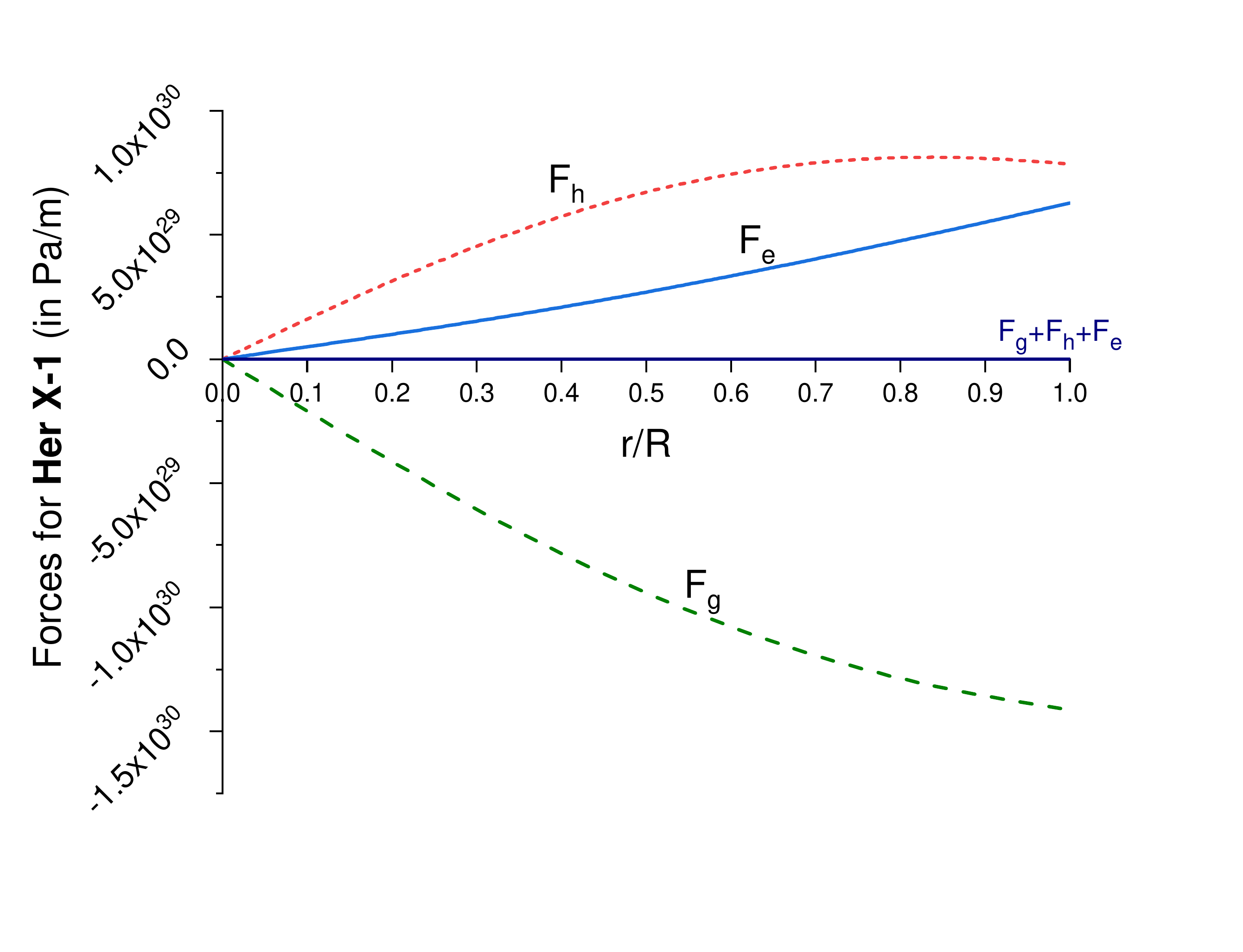}\includegraphics[width=8cm]{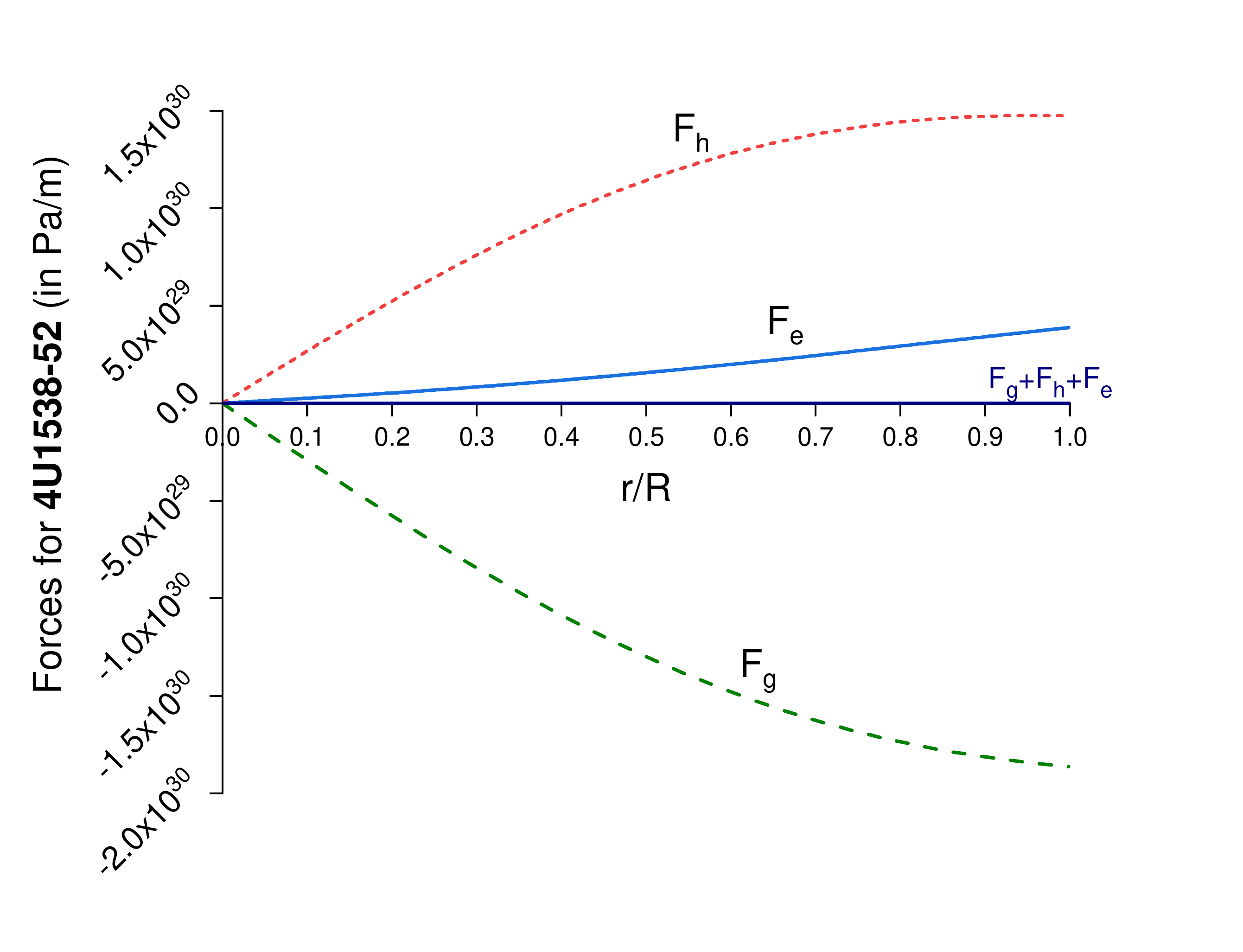}
						\includegraphics[width=8cm]{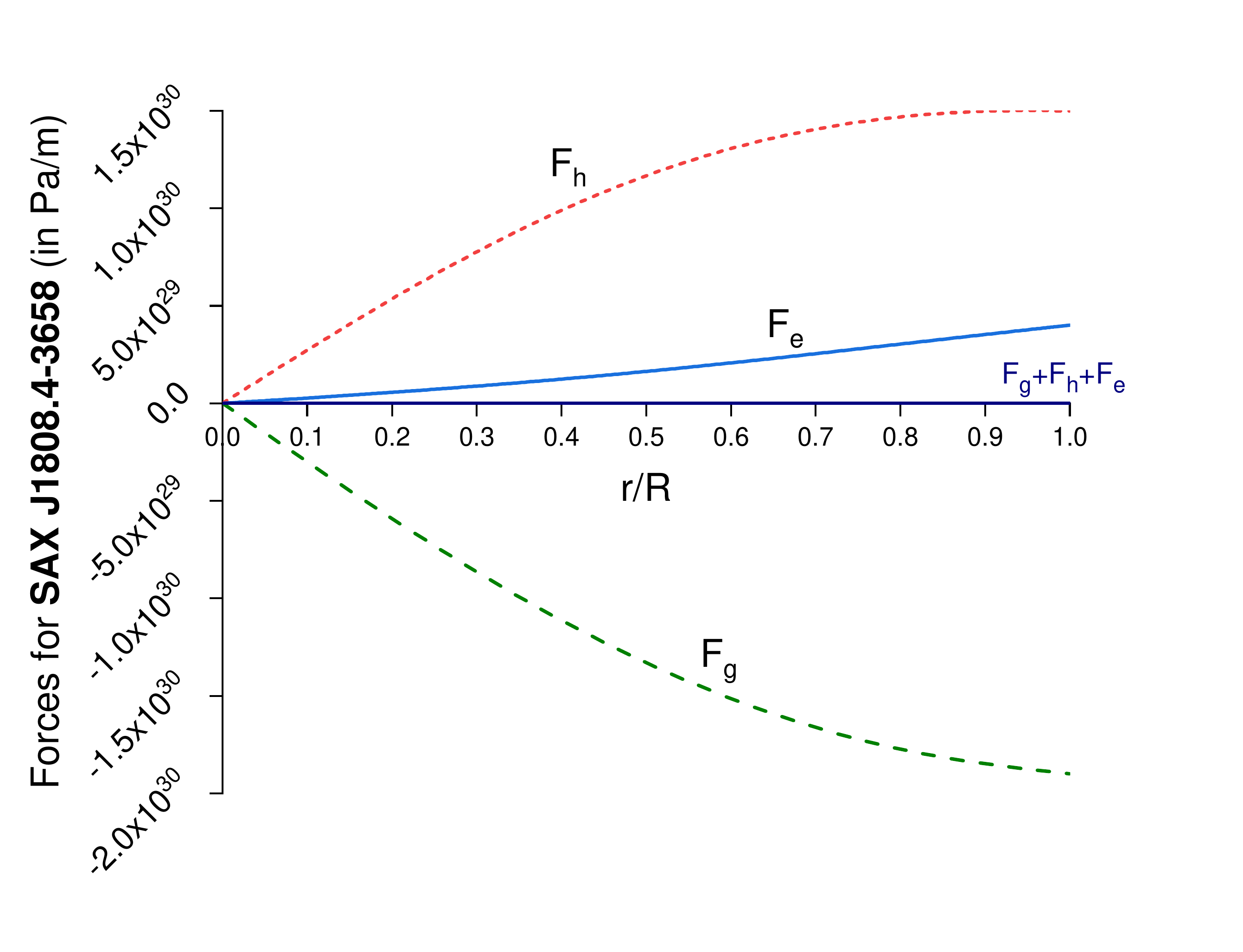}\includegraphics[width=8cm]{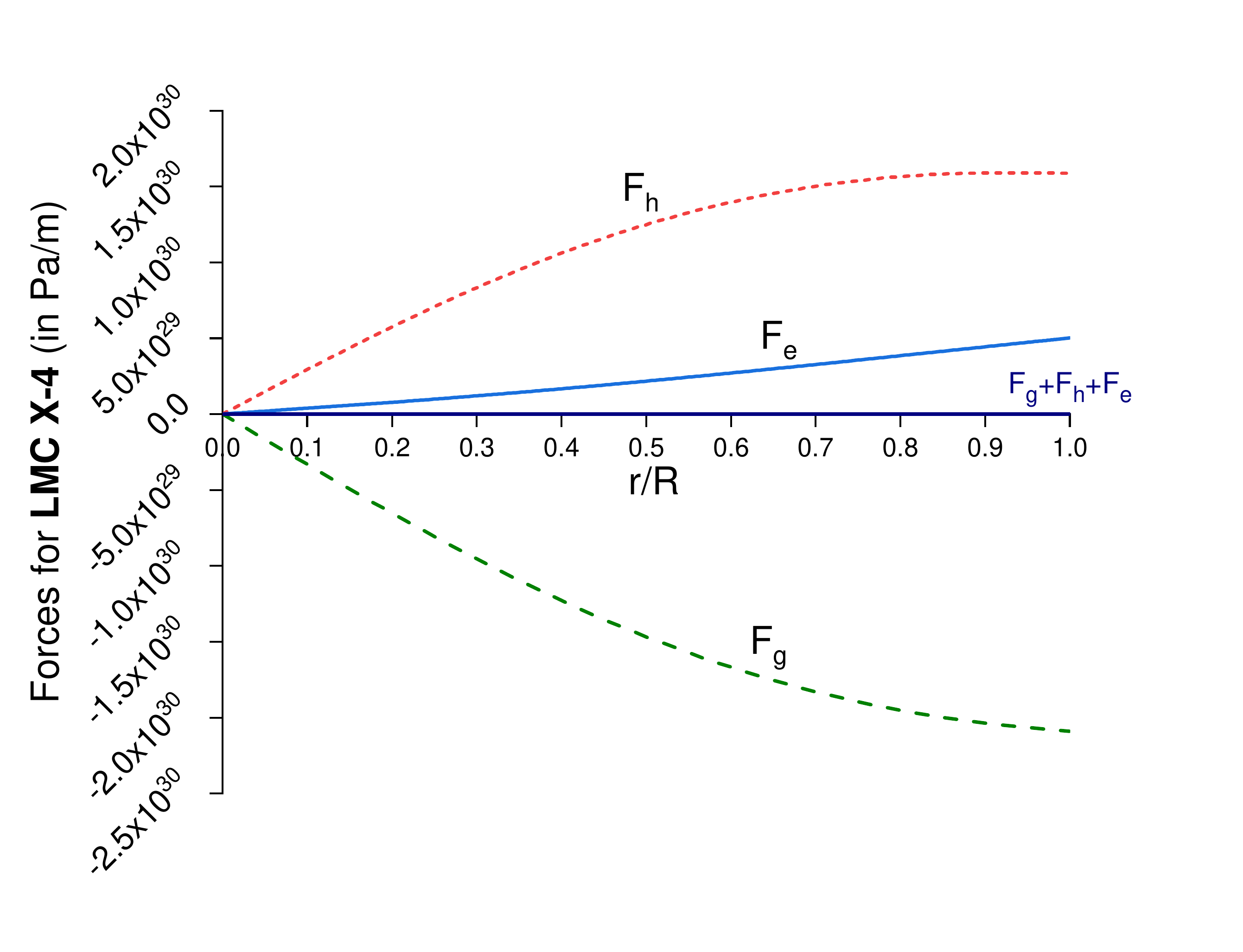}
						\includegraphics[width=8cm]{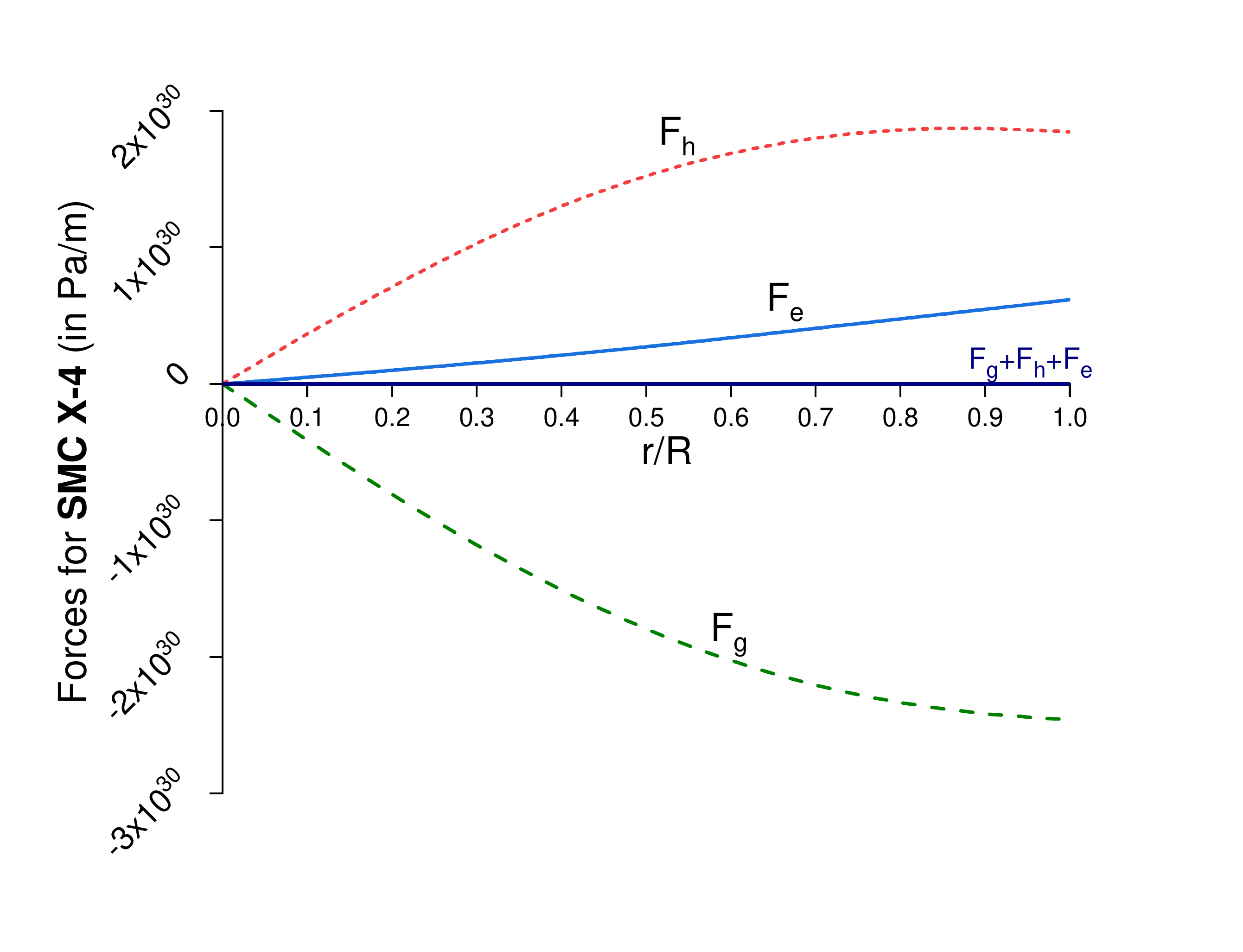}\includegraphics[width=8cm]{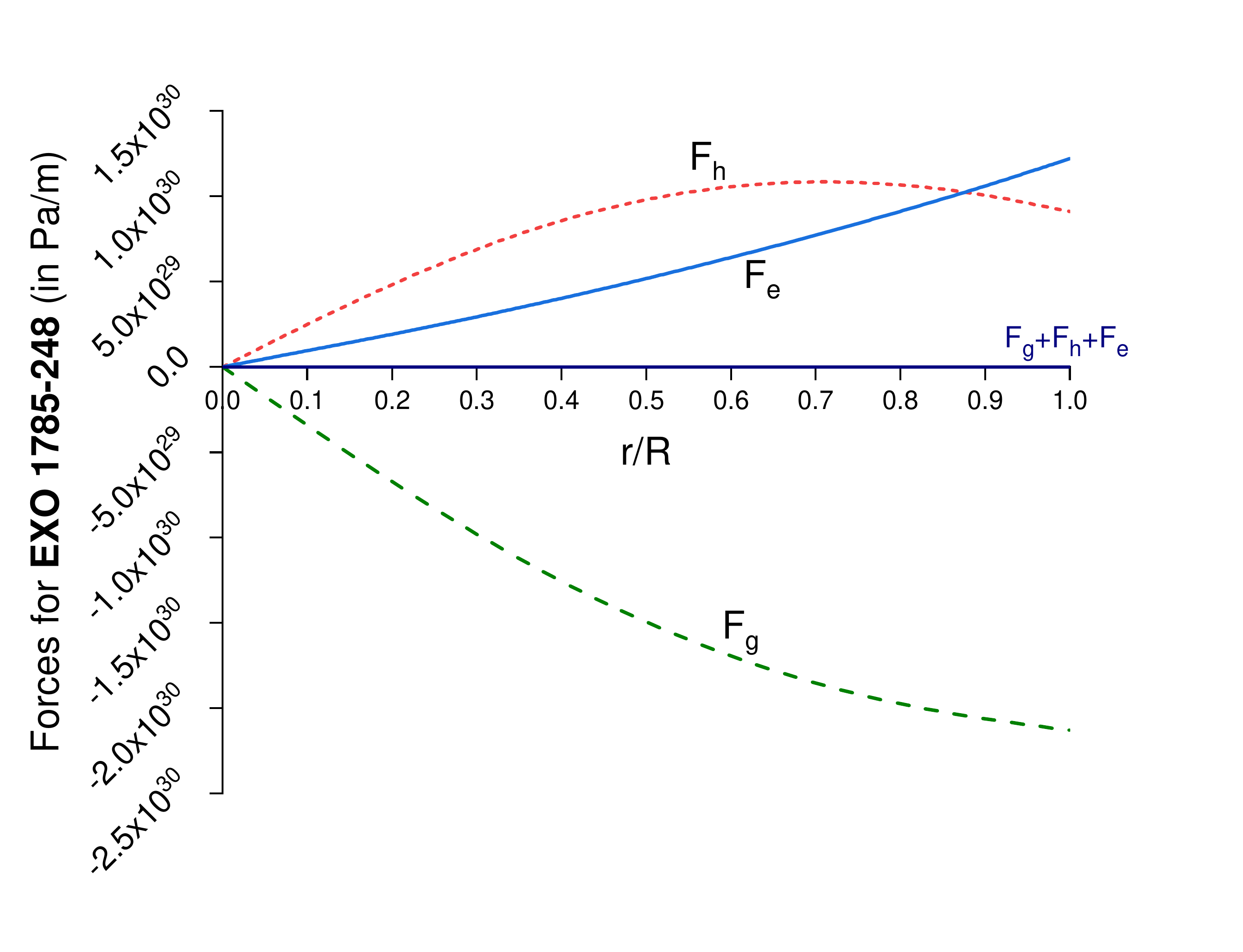}
						\includegraphics[width=8cm]{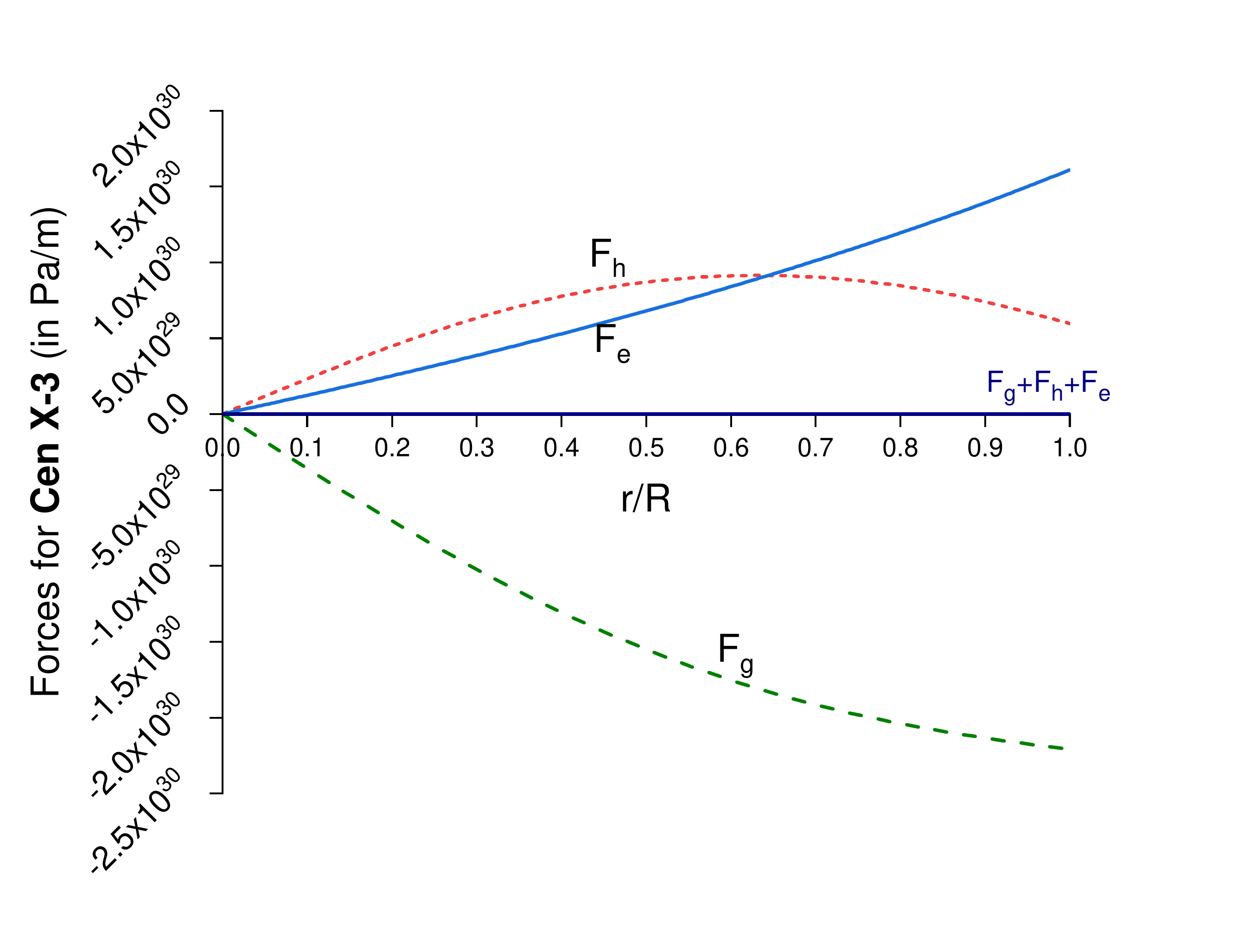}\includegraphics[width=8cm]{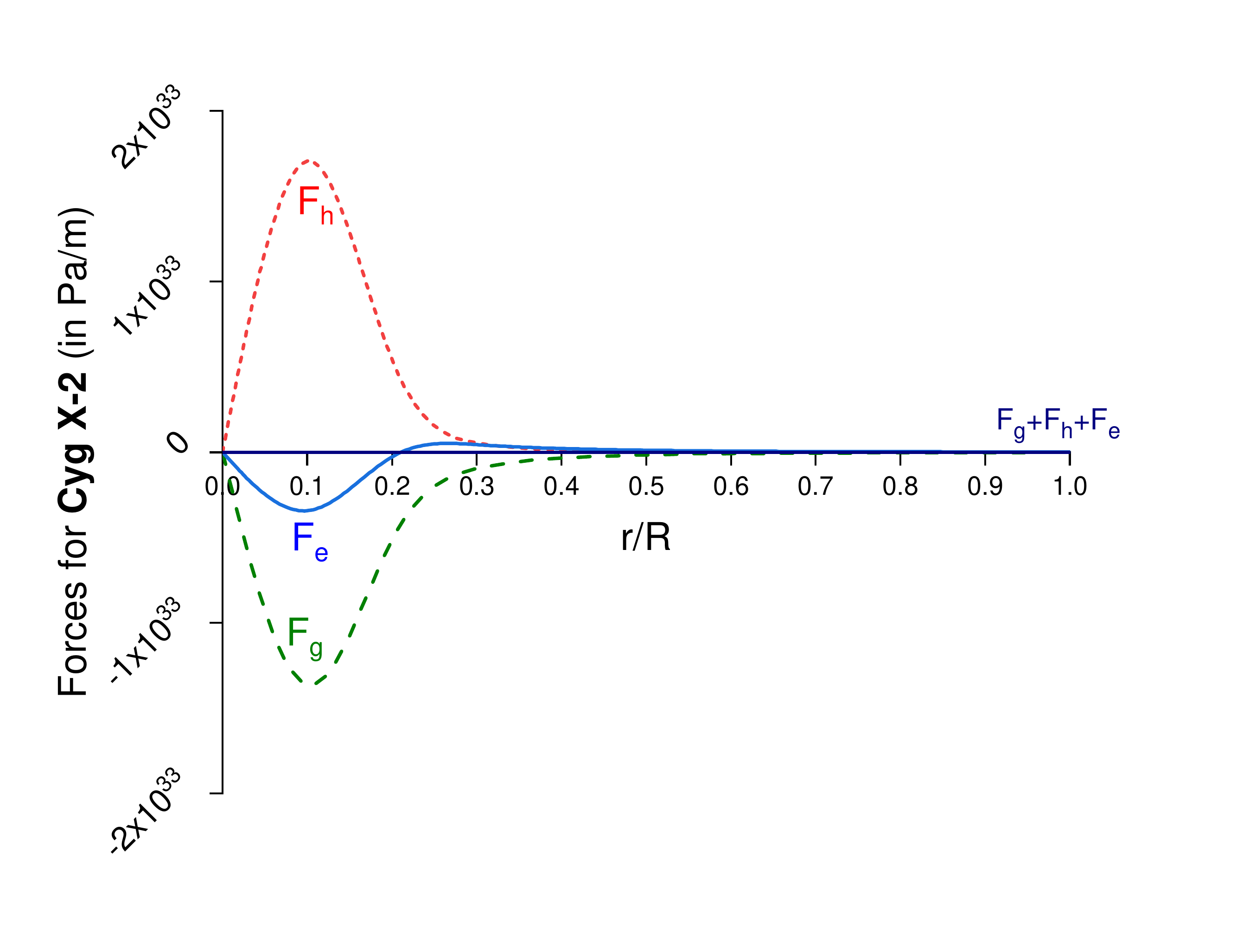}
						\caption{Variations of gravitational force ($F_{g}$),  hydrostatic force ($F_{h}$)  and electric force ($F_{e}$) acting on the system with respect to fractional radius ($r/R$) for star candidates Her X-1, 4U 1538-52, SAX J1808.4-3658, LMC X-4, SMC X-4,  EXO 1785-248, Cen X-3  ($K<0$) and Cyg X-2 ($K>1$).} \label{t}
					\end{figure}

					We have drawn figures  for each compact star candidates to show the behaviour of these forces. It is evident from fig (\ref{t}) that $F_{g}$ nullifies the combined effect of $F_{h}$ and $F_{e}$. In other words,  the static equilibrium is attainable under these three different forces for this model.
					\subsection{\textbf{Relativistic Adiabatic Index}}
					The adiabatic index  
					\begin{equation}
						\gamma=\Big(\frac{c^2\rho+p}{p} \Big)\Big(\frac{dp}{c^2 d\rho}\Big)
					\end{equation}
					is related to the stability of a stellar configuration. For an isotropic star to be in stable equilibrium, $\gamma$ must have values strictly greater than $\frac{4}{3}$ throughout the region.
					\begin{figure}[H]
							\includegraphics[width=8cm]{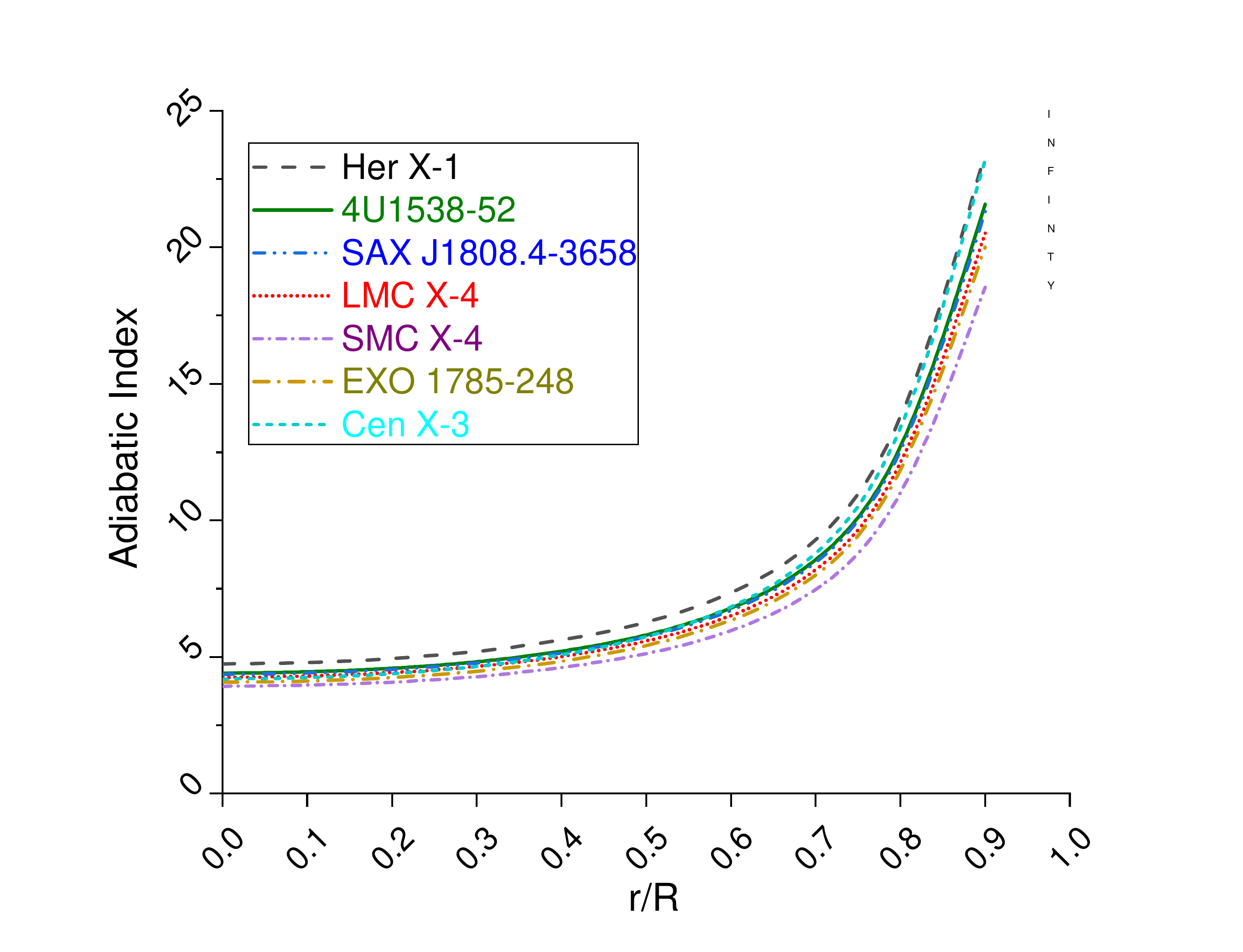}\includegraphics[width=8cm]{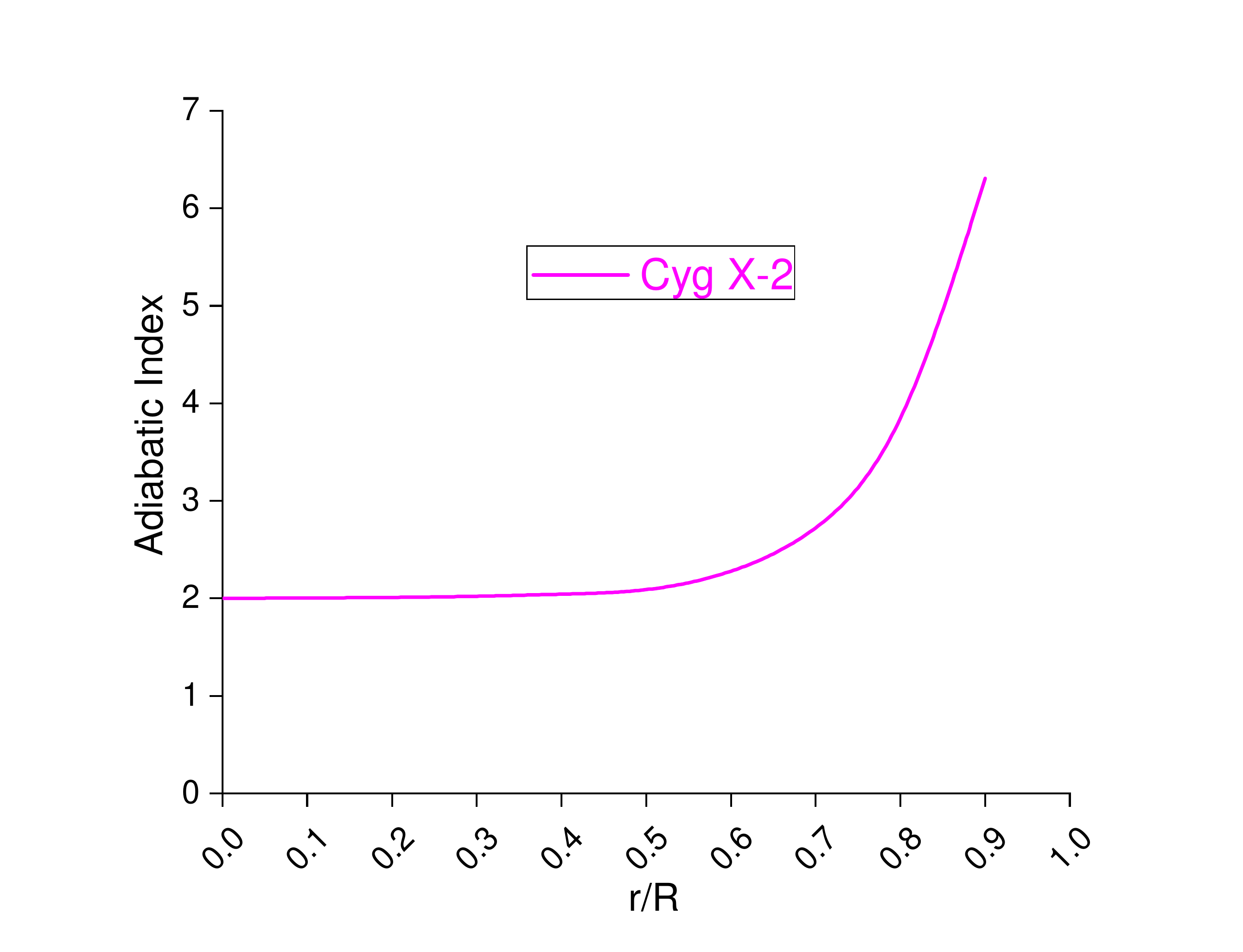}
							\caption{Variation of adiabatic index with respect to fractional radius (r/R) for star candidates Her X-1, 4U 1538-52, SAX J1808.4-3658, LMC X-4, SMC X-4,  EXO 1785-248, Cen X-3  ($K<0$) and Cyg X-2 ($K>1$).}\label{adia}
					\end{figure}
					Graphs in fig. (\ref{adia}) represent the behavior of adiabatic index $\gamma$. We can see that the desirable features have been obtained for each star candidate that we have considered.
					\subsection{\textbf{Harrison-Zeldovich-Novikov Stability Criterion}}
					Harrison-Zeldovich-Novikov criterion \cite{harrison, zeldovich} states the condition for stability of a compact object. According to this criterion, to have a stable configuration, mass of a compact star should increase with increase in central density throughout the stellar region. Mathematically, $\frac{dM}{d\rho_0}>0$  \\
					\begin{figure}[h]
						\begin{center}
							\includegraphics[width=8cm]{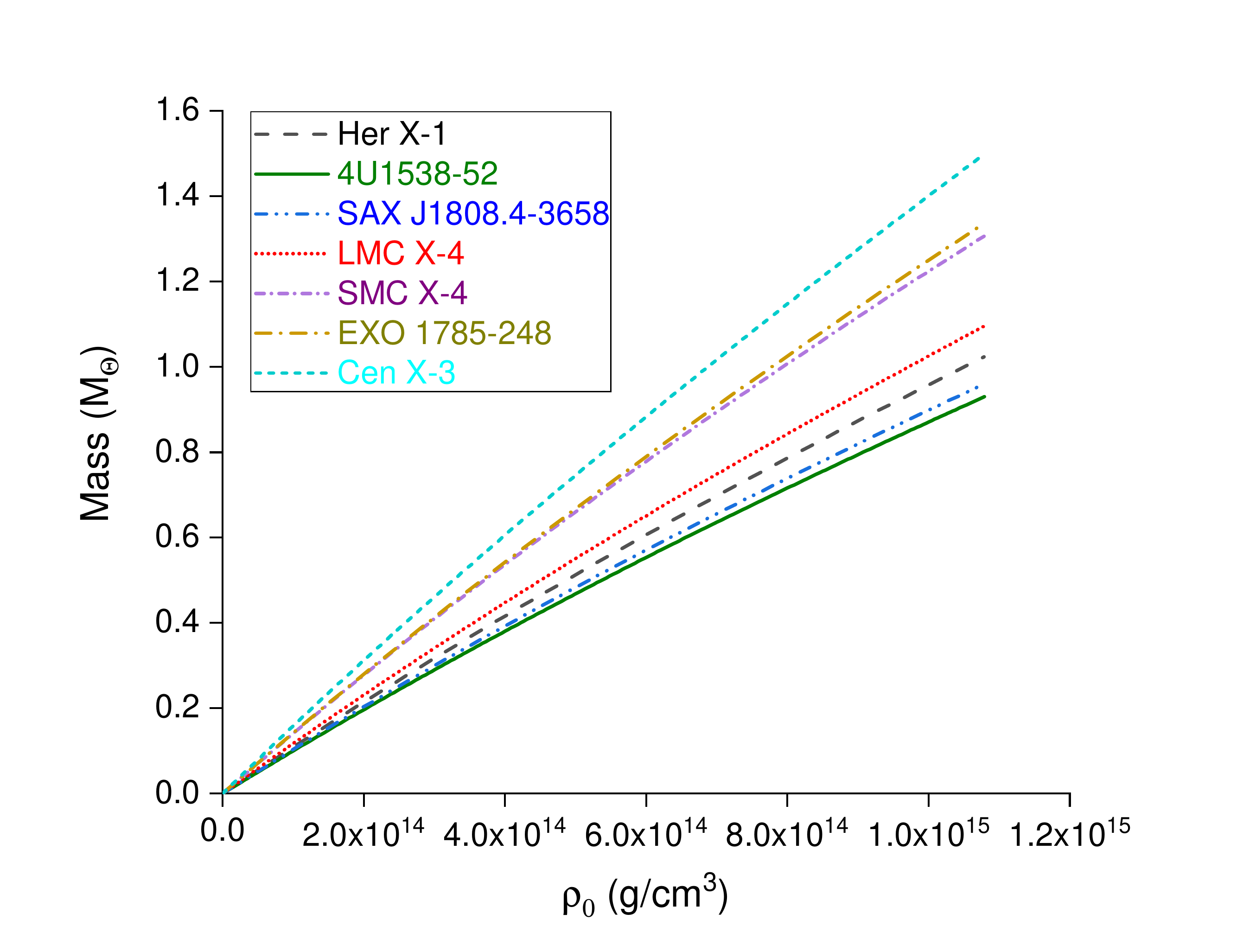}\includegraphics[width=8cm]{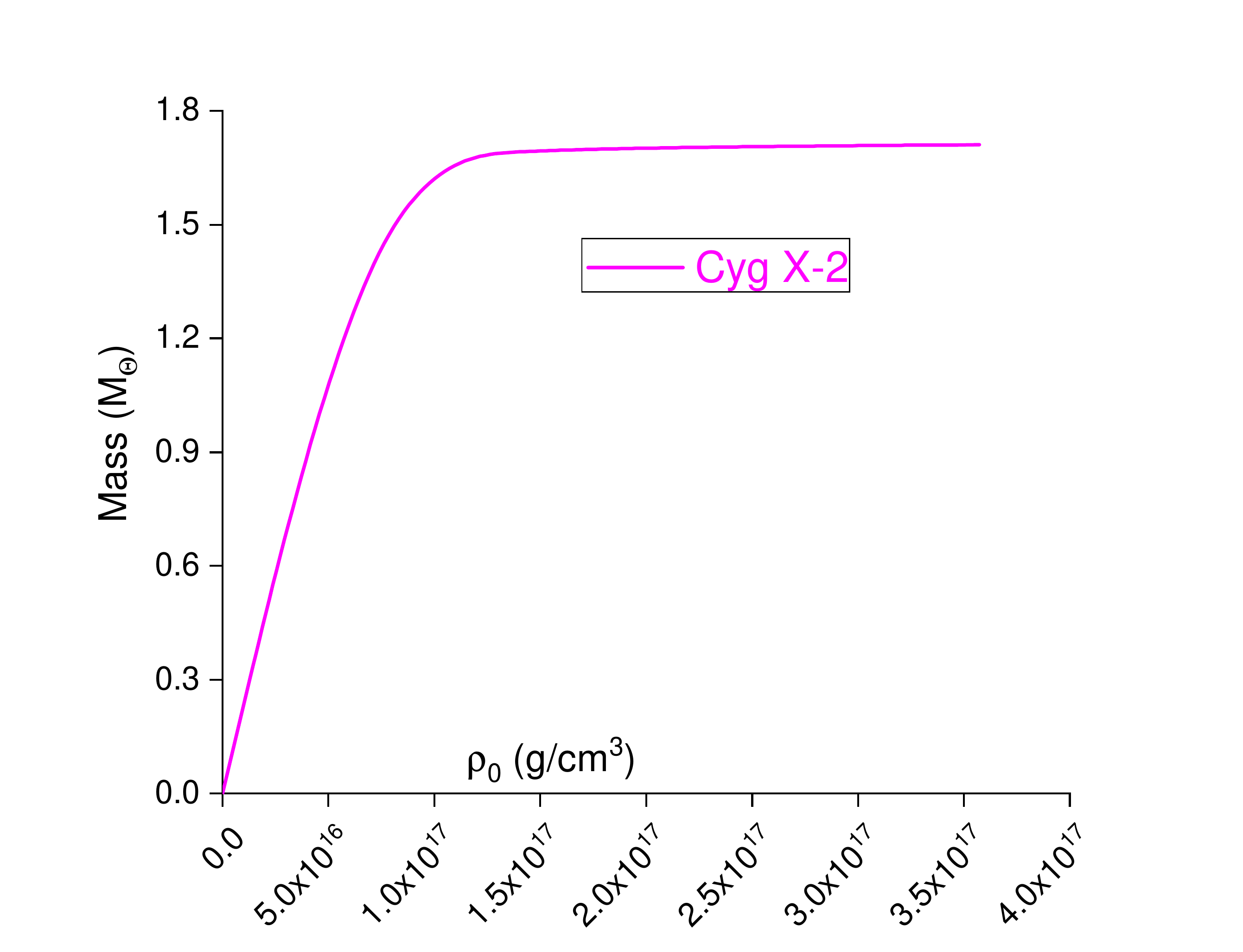}
							\caption{Variation of mass function with respect to density for star candidates Her X-1, 4U 1538-52, SAX J1808.4-3658, LMC X-4, SMC X-4,  EXO 1785-248, Cen X-3  ($K<0$) and Cyg X-2 ($K>1$).}\label{m}
						\end{center}
					\end{figure}
					We can obtain the expression for central density from eq. (\ref{density}) as
					\begin{equation}
						\rho_0=\frac{3C(K-1)}{8\pi K}
						\label{centdens}
					\end{equation}
					Let's write $M$ in terms of $\rho_0$ using eqs. (\ref{bcon1}) and (\ref{centdens}) as 
					\begin{equation}
						M=4\pi R^3\frac{\rho_0}{M_1}  \Big[K-1+4\pi KR^2\rho_0  \frac{ M_2}{M_1}\Big]
					\end{equation}
					where,\\
					$M_1=3(K-1)+8\pi K\rho_0 R^2$,\\
					$M_2=\Big[-\frac{15}{4} \frac{(1-K)^2}{M_1}-\frac{2a_1M_1}{3(1-K)^2X_1^2(a_1+a_2X_1)}+K-\frac{7}{4}\Big]$.\\
					
					We can observe in fig (\ref{m}) that mass of every star is positive definite and it increases with  increase in central density. Thus, we can conclude that the presented model satisfies Harrison-Zeldovich-Novikov criterion of stability. 
					
					\section{Conclusion}
					In this paper, we have investigated the nature of isotropic compact stars. By employing the Vaidya and Tikekar ansatz for metric potential, we have simplified the Einstein field equations and obtained exact solutions for isotropic compact stars. Based on physical requirements, we equated the interior solution to the exterior one (Reissner-N\"{o}rdstro solution) at surface to fix the value of constants $A_1$ and $A_2$ . Using these values of constants and free parameters $ C $ and $ K $ it is possible to determine mass and radius for compact stars. To refine the model further, we have shown through graphs, that metric potentials are regular, energy density and pressure are finite at the center and monotonically decreasing towards the boundary. The pressure vanishes at the boundary.
					The electric field intensity is zero at centre and it increases towards the surface.
					We have shown that the model is compatible with the compact objects such as, Her X-1, 4U 1538-52, SAX J1808.4-3658, LMC X-4, SMC X-4,  EXO 1785-248, Cen X-3  and Cyg X-2. As shown in Table \ref{t2}, the gravitational redshift is bounded above function and satisfies $z_s < 2$. 
					Adiabatic index is strictly greater than $\frac{4}{3}$ throughout the model. The model satisfies the TOV equation, energy conditions, the causality condition and it also fullfills Harrison-Zeldovich-Novikov criterion. This  shows that the obtained model is stable.
					To obtain numerical values of physical quantities, we have taken $G=6.674 \times 10^{-11}N/ms^2, \ c=3\times 10^8 m/s, \ 1M_\odot=1.475km$ and have multiplied charge by $1.1659\times 10^{20}$ to convert it from relativistic unit ($km$) to  coulomb.
					
					As a future scope, we can look for other forms of metric potentials which could possess more general behaviour and thus it might be able to describe other types of compact objects. 
					\section*{Acknowledgments}
					The Authors are sincerely grateful towards Science and Engineering Research Board (SERB), DST, New Delhi for providing the needed financial support. They are also very humbled towards the Department of Mathematics, Central University of Jharkhand, Ranchi, India, where the paper has been written and finalized, for the much needed support.
					
					\section*{Appendix A: Notations used in eqs (\ref{pressure}), (\ref{grad.d}) and (\ref{grad.p})}
					$P_1= \frac{2}{(K-1)X}\Big(\frac{a_1+a_2X}{2(X^2-1)}-\frac{a_1}{X^2}\Big)$, \ 
					$P_2=\frac{a_1}{a_2^3}S(X)+\frac{A_2}{A_1}$, \ 
					$P_3=\frac{a_1}{a_2^2(K-1)X^2}$, \
					$P_4=\sec^2\Big(\tan^{-1}\sqrt{\frac{a_2X}{a_1}}\Big)+\cos^2\Big(\tan^{-1}\sqrt{\frac{a_2X}{a_1}}\Big)-2$, \
					$P_5=\Big(\frac{a_1+a_2X}{X}\Big)$ \\
					$ D(r)=D_1 P_2+ D_2P_1+D_3 P_4+D_4 P_3-\Big(\frac{P_1 P_2+P_3 P_4}{P_2 P_5}\Big) \Big(D_2 P_5+D_5 P_2\Big)$, \\
					$D_1(r)=-\frac{P_1}{(K-1)X^2}+\frac{2}{(K-1)^2X^2}\Big[\frac{a_2}{2(X^2-1)}-\frac{X(a_1+a_2X)}{(X^2-1)^2}+\frac{2a_1}{X^3}\Big]$, \
					$D_2(r)=\frac{a_1}{2a_2^2(K-1)}\frac{P_4}{X(a_1+a_2X)}$, \ 
					$D_3(r)=\frac{-2a_1}{(K-1)^2a_2^2X^4}$, \
					$D_4(r)=\frac{a_2}{(K-1)X(a_1+a_2X)}\Big\{\sec^2\Big(\tan^{-1}\sqrt{\frac{a_2X}{a_1}}\Big)-\cos^2\Big(\tan^{-1}\sqrt{\frac{a_2X}{a_1}} \Big)\Big\}$, \
					$D_5(r)=\frac{-a_1}{(K-1)X^3}$, \
					$D_6(r)=\frac{2(1-K)(5+Cr^2)}{K(1+Cr^2)^3}$, \
					$D_7(r)=\frac{1-Cr^2}{K(1+Cr^2)^3}\Big[\frac{5}{4(1-X^2)}-\frac{2a_1(1-X^2)}{X^2(a_1+a_2X)}+K-\frac{7}{4}\Big]$, \
					$D_8(r)=\frac{Cr^2}{4K(K-1)(1+Cr^2)^2}\Big[\frac{-5}{(1-X^2)^2}+\frac{8a_1}{X^2(a_1+a_2X)}+\frac{4a_1(1-X^2)(2a_1+3a_2X)}{X^4(a_1+a_2X)^2}\Big]$.
					\section*{Appendix B : Finding constants $A_1$ \& $A_2$}
					Here we are going to calculate the values of arbitrary constants $A_1$ and $A_2$, used in eq. (\ref{solz}), using boundary conditions (\ref{bcon1},\ref{bcon2}).\\
					First we are going to determine the value of $\frac{A_2}{A_1}$. Using boundary conditon $p(R)=0$ in eq. (\ref{pressure}), we can obtain the following relationship
					\begin{equation}
						-\frac{P_{31}  P_{41}   (K+CR^2)}{P_{11} (K+CR^2)+(1-K) (J_1+1)  P_{51}}=P_{21}=\frac{a_1}{a_2^3}S(X_1)+\frac{A_2}{A_1}
					\end{equation}
					Thus, we have
					\begin{equation}
						\frac{A_2}{A_1}=\Big[\frac{P_{31}P_{41} X_1^2}{(J_1+1) P_{51}-P_{11} X_1^2}-\frac{a_1}{a_2^3}S(X_1)\Big]
						\label{A_2/A_1}
					\end{equation}
					where,\\
					$X_1=\sqrt{\frac{K+CR^2}{K-1}}$, \
					$J_1=\frac{CR^2(1-X_1^2)}{2(1+CR^2)^2}\Big[\frac{5}{4(1-X_1)}-\frac{2a_1(1-{X_1}^2)}{{X_1}^2(a_1+a_2X_1)}+K-\frac{7}{4}\Big]$, \ 
					$P_{11}= \frac{2}{(K-1)X_1}\Big(\frac{a_1+a_2X_1}{2(X_1^2-1)}-\frac{a_1}{X_1^2}\Big)$, \ 
					$P_{31}=\frac{a_1}{a_2^2(K-1)X_1^2}$, \ 
					$P_{41}=\sec^2\Big(\tan^{-1}\sqrt{\frac{a_2X_1}{a_1}}\Big)+\cos^2\Big(\tan^{-1}\sqrt{\frac{a_2X_1}{a_1}}\Big)-2$, \ 
					$\&$ \ $P_{51}=\Big(\frac{a_1+a_2X_1}{X_1}\Big)$. \\ \\ 
					To find the value of $A_1$, we will use the condition $Z^2(R)=\frac{K+CR^2}{K(1+CR^2)}$. After a little bit of computation, we can obtain the values of $A_1$ and $A_2$ as,
					\begin{eqnarray}
						\nonumber A_1=\frac{1}{\sqrt{K}(1-{X_1}^2)^{3/4}} \Big [\frac{(J_1+1)  P_{51}-P_{11}  X_1^2}{(a+bX_1)  P_{31}  P_{41}}\Big], \ \ when \ \ K<0\\
						A_1=\frac{1}{\sqrt{K}({X_1}^2-1)^{3/4}} \Big [\frac{(J_1+1)  P_{51}-P_{11}  X_1^2}{(a+bX_1)  P_{31}  P_{41}}\Big],  \ \ when \ \ K>1
						\label{A}
					\end{eqnarray}
					and 
					\begin{eqnarray}
						\nonumber A_2=\frac{1}{\sqrt{K}(1-{X_1}^2)^{3/4}} \Big[\frac{X_1^2}{a_1+a_2X_1}-\frac{a_1}{a_2^3}\frac{(J_1+1)  (P_{51}-P_{11}  X_1^2)}{(a_1+a_2X_1)  P_{31}  P_{41}}S(X_1)\Big], \ \  when \ \ K<0 \\
						A_2=\frac{1}{\sqrt{K}({X_1}^2-1)^{3/4}} \Big[\frac{X_1^2}{a_1+a_2X_1}-\frac{a_1}{a_2^3}\frac{(J_1+1)  (P_{51}-P_{11}  X_1^2)}{(a_1+a_2X_1)  P_{31}  P_{41}}S(X_1)\Big], \ \ when \ \ K>1. \ \ \
						\label{B}
					\end{eqnarray}

					\section*{Appendix C: Structural properties of compact stars in relativistic units}
					\begin{table}[h!]
						\caption{Structural properties of  \textquotedblleft Her X-1\textquotedblright within radius.}
						\begin{tabular}{cccccccc}
							\toprule
							$r/R$ & $q(km)$ &$\rho(km^{-2})$&$\sigma (km^{-2})$&$p (km^{-2})$& $p/\rho$ & $v_s^2$&$\gamma$\\
							\midrule
							$0$&     $0$     &$0.000647295$&$0.000343118$&$3.51166\times 10^{-5}$&$0.054246$&$0.243597$&$4.73422$\\
							$0.2$&$0.0062208$&$0.000638455$&$0.000341655$&$3.29828\times 10^{-5}$&$0.051653$&$0.241651$&$4.919966$\\
							$0.4$&$0.0519939$&$0.000612955$&$0.000337174$&$2.70386\times 10^{-5}$&$0.044117$&$0.235845$&$5.58173$\\
							$0.6$&$0.1854576$&$0.000573693$&$0.000329066$&$1.85338\times 10^{-5}$&$0.032308$&$0.226207$&$7.227869$\\
							$0.8$&$0.4631823$&$0.0005246$&$0.000316964$&$9.02301\times 10^{-6}$&$0.017206$&$0.212543$&$12.565491$\\
							$0.1$&$0.9437229$&$0.000469944$&$0.000301478$&$0$&$0$&$0.194044$&$Inf$\\
							\bottomrule							
						\end{tabular}
						\label{t4}
					\end{table}
					\begin{table}[h!]
						\caption{Structural properties of \textquotedblleft 4U 1538-52\textquotedblright within radius.}
						\begin{tabular}{cccccccc}
								\toprule
							$r/R$ & $q(km)$ &$\rho(km^{-2})$&$\sigma (km^{-2})$&$p (km^{-2})$& $p/\rho$ & $v_s^2$&$\gamma$\\
							\midrule
							$0$&$0$&$0.000740376$&$0.000250639$&$6.09788\times 10^{-5}$&$0.082368$&$0.335101$&$4.403433$\\
							$0.2$&$0.004255506$&$0.000730016$&$0.000249992$&$5.75363\times 10^{-5}$&$0.078808$&$0.333976$&$4.571807$\\
							$0.4$&$0.037426428$&$0.000700214$&$0.000248053$&$4.7823\times 10^{-5}$&$0.068302$&$0.330724$&$5.172831$\\
							$0.6$&$0.140156388$&$0.000654508$&$0.000243333$&$3.35844\times 10^{-5}$&$0.051301$&$0.325676$&$6.673971$\\
							$0.8$&$0.362087712$&$0.00059778$&$0.000233507$&$1.69538\times 10^{-5}$&$0.028372$&$0.319259$&$11.572042$\\
							$0.1$&$0.750605184$&$0.000535233$&$0.000217701$&$0$&$0$&$0.311849$&$Inf$\\
							\bottomrule
						\end{tabular}
						\label{t5}
					\end{table}
					\begin{table}
						\caption{Structural properties of \textquotedblleft SAX J1808.4-3658\textquotedblright within radius.}
						\begin{tabular}{cccccccc}
								\midrule
							$r/R$ & $q(km)$ &$\rho(km^{-2})$&$\sigma (km^{-2})$&$p (km^{-2})$& $p/\rho$ & $v_s^2$&$\gamma$\\
							\midrule
							$0$&0&0.000742648&0.000256223&$6.28298\times 10^{-5}$&0.084607&0.339965&4.358118\\
							$0.2$&0.004484364&0.000732097&0.000255369&$5.92707\times 10^{-5}$&0.080957&0.338846&4.52437\\
							$0.4$&0.039317695&0.000701789&0.000252743&$4.92578\times 10^{-5}$&0.070181&0.335617&5.117791\\
							$0.6$&0.146719803&0.000655363&0.000247017&$3.45627\times 10^{-5}$&0.052735&0.330607&6.599795\\
							$0.8$&0.377966687&0.000597816&0.000236071&$1.74475\times 10^{-5}$&0.029183&0.324236&11.434827\\
							$0.1$&0.781734369&0.00053448&0.000219208&$0$&0&0.316856&$Inf$\\						
								\bottomrule
						\end{tabular}
						\label{t6}
					\end{table}
					\begin{table}[h!]
						\caption{Structural properties of \textquotedblleft LMC X-4\textquotedblright within radius.}
						\begin{tabular}{cccccccc}
							\midrule
							$r/R$ & $q(km)$ &$\rho(km^{-2})$&$\sigma (km^{-2})$&$p (km^{-2})$& $p/\rho$ & $v_s^2$&$\gamma$\\
						\midrule
						$0$&0&0.000753411&0.000299159&$7.02981\times 10^{-5}$&0.093311&0.362564&4.248113\\
						$0.2$&0.00589371&0.000742309&0.000296924&$6.63071\times 10^{-5}$&0.089316&0.361449&4.40832\\
						$0.4$&0.05038707&0.000710469&0.000290393&$5.506\times 10^{-5}$&0.077505&0.358201&4.979845\\
						$0.6$&0.183568314&0.000661794&0.000279131&$3.8603\times 10^{-5}$&0.05833&0.353044&6.405616\\
						$0.8$&0.464499057&0.000601669&0.000262529&$1.94611\times 10^{-5}$&0.032339&0.34615&11.049986\\
						$0.1$&0.950041149&0.000535754&0.000241138&$0$&0&0.337298&$Inf$\\
							\bottomrule
						\end{tabular}
						\label{t7}
					\end{table}
					\begin{table}[h!]
					\centering
					\caption{Structural properties of  \textquotedblleft SMC X-4\textquotedblright within radius.}
					\begin{tabular}{cccccccc}
						\midrule
						$r/R$ & $q(km)$ &$\rho(km^{-2})$&$\sigma (km^{-2})$&$p (km^{-2})$& $p/\rho$ & $v_s^2$&$\gamma$\\
						\midrule
						$0$&0&0.00078779&0.000325954&$9.00668\times 10^{-5}$&$0.114335$&0.402212&3.920058\\
						$0.2$&0.007727125&0.000774608&0.000322645&$8.47967\times 10^{-5}$&$0.109471$&0.401061&4.064701\\
						$0.4$&0.065923415&0.00073696&0.00031308&$7.00762\times 10^{-5}$&$0.095082$&0.397692&4.580329\\
						$0.6$&0.239090494&0.000680079&0.000297205&$4.8752\times 10^{-5}$&$0.071677$&0.392261&5.864862\\
						$0.8$&0.601355776&0.000610913&0.00027515&$2.43375\times 10^{-5}$&$0.039836$&0.384667&10.040912\\
						$0.1$&1.222634288&0.000536426&0.000248761&0&$0$&0.373824&$Inf$\\	
						\bottomrule
					\end{tabular}
					\label{t8}
				\end{table}
			\begin{table}[h!]
				\centering
				\caption{Structural properties of  \textquotedblleft EXO 1785-248 \textquotedblright within radius.}
				\begin{tabular}{cccccccc}
					\midrule
					$r/R$ & $q(km)$ &$\rho(km^{-2})$&$\sigma (km^{-2})$&$p (km^{-2})$& $p/\rho$ & $v_s^2$&$\gamma$\\
					\midrule
					$0$&0&0.000774921&0.000455298&$5.20913\times 10^{-5}$&0.067224&0.256364&4.069953\\
					$0.2$&0.010760384&0.00076072&0.00045185&$4.849\times 10^{-5}$&0.063746&0.253618&4.232158\\
					$0.4$&0.089843897&0.000720288&0.000441454&$3.87205\times 10^{-5}$&0.053752&0.245302&4.808875\\
					$0.6$&0.319227675&0.000659461&0.000424342&$2.53624\times 10^{-5}$&0.03846&0.230946&6.2358\\
					$0.8$&0.792950041&0.00058589&0.000401993&$1.15446\times 10^{-5}$&0.019701&0.208868&10.810668\\
					$0.1$&1.609960513&0.000506878&0.000378189&$0$&0&0.174329&$Inf$\\	
					\bottomrule
				\end{tabular}
				\label{t9}
			\end{table}
				\begin{table}[h!]
					\centering
					\caption{Structural properties of  \textquotedblleft Cen X-3 \textquotedblright within radius.}
					\begin{tabular}{cccccccc}
						\midrule
						$r/R$ & $q(km)$ &$\rho(km^{-2})$&$\sigma (km^{-2})$&$p (km^{-2})$& $p/\rho$ & $v_s^2$&$\gamma$\\
					\midrule
					$0$&0&0.000794746&0.000514889&$4.36632\times 10^{-5}$&0.054945&0.218321&4.191785\\
					$0.2$&0.013555906&0.000778471&0.000510378&$4.0173\times 10^{-5}$&0.051602&0.214795&4.377354\\
					$0.4$&0.11279762&0.000732397&0.000496939&$3.09014\times 10^{-5}$&0.042185&0.204072&5.041574\\
					$0.6$&0.398940126&0.00066378&0.000475404&$1.88637\times 10^{-5}$&0.028424&0.185362&6.706768\\
					$0.8$&0.98686445&0.000581808&0.000448432&$7.57399\times 10^{-6}$&0.013016&0.156026&12.14358\\
					$0.1$&2.000262498&0.00049485&0.000421627&$0$&0&0.108787&$Inf$\\	
							\bottomrule
						\end{tabular}
						\label{t10}
					\end{table}
					\begin{table}[h!]
						\caption{Structural properties of  \textquotedblleft Cyg X-2 \textquotedblright within radius.}
						\begin{tabular}{cccccccc}
							\midrule
							$r/R$ & $q(km)$ &$\rho (km^{-2})$&$\sigma (km^{-2})$ &$p (km^{-2})$& $p/\rho$ & $v_s^2$&$\gamma$\\
							\midrule
							$0$  & $0$         &0.264256351&0.246886517&0.07718148          &0.292071 & 0.513242&2.000395\\
							$0.2$&0.37743189&0.007328578&0.001769387&$0.000466881$&0.063707& 0.158856&2.008239\\
							$0.4$& 0.91447395&0.001657421&0.000655655&$7.66333\times 10^{-5}$& 0.046239&0.090316&2.04356\\
							$0.6$&1.417686&0.000719241&0.000307866&$2.23442\times 10^{-5}$&0.031063& 0.067565& 2.242633\\
							$0.8$&1.91133324 &0.000401051&0.000176378&$6.28475\times 10^{-6}$&0.015668 &0.054466&3.530793\\
							$1.0$& 2.40103323&0.000255633&0.000113835&$0$                 &0          & 0.044416 &$Inf.$\\
							\bottomrule
						\end{tabular}
						\label{t11}
					\end{table}
					
				\end{document}